\documentclass{tlp}

\usepackage{pifont}
\usepackage{amsmath}
\usepackage[dvips]{graphicx}
\usepackage{calc}
\usepackage{fancyvrb}
\usepackage{pstricks}
\usepackage[matrix,arrow,frame]{xy}
\usepackage{makeidx}
\usepackage{epsfig}
\usepackage{subfigure}
\newcounter{cncounter}

\newcommand{\ntdom}{\ensuremath{L_{\texttt{notrail}}}}
\newcommand{\tdom}{\ensuremath{L_{\texttt{trail}}}}

\newenvironment{example}{\begin{ex}\rm}{\end{ex}}
\newtheorem{ex}{Example}

\newcounter{xmpl}[section]

\newenvironment{cxmpl}[1]{\bgroup
\par\noindent\it Counterexample.\it\ 
\rm
#1}{\par\indent\egroup}

\begin{document}

\title[Improving PARMA trailing]{Improving PARMA trailing}
\author[T. Schrijvers et al.]{TOM SCHRIJVERS\thanks{Research Assistant of the fund for Scientific Research - Flanders (Belgium)(F.W.O. - Vlaanderen)} and BART DEMOEN \\
       Dept. of Computer Science, K.U.Leuven, Belgium \\
       \email{\{toms,bmd\}@cs.kuleuven.ac.be}
       \and
       MARIA GARCIA DE LA BANDA\\
       School of Computer Science and S.E, Monash University, Australia \\
       \email{mbanda@csse.monash.edu.au}
       \and
       PETER J. STUCKEY \\
       NICTA Victoria Laboratory\\
       Department of Computer Science and S.E. \\
       University of Melbourne, Australia
       \\
       \email{pjs@cs.mu.oz.au}
}
\submitted{20 November 2003}
  \revised{22 October 2004}
  \accepted{30 May 2005}
\maketitle

\begin{abstract}
Taylor introduced a variable binding scheme for logic variables
in his PARMA system, that uses cycles of bindings rather than 
the linear chains of bindings used in the standard WAM representation. 
Both the HAL and dProlog languages make use of the PARMA 
representation
in their Herbrand constraint solvers.
Unfortunately, PARMA's
trailing scheme is considerably more expensive in both time and
space consumption. The aim of this paper is to 
present several techniques that lower the cost.

First, we introduce a trailing analysis for HAL using 
the classic PARMA trailing scheme that detects and eliminates unnecessary
trailings. The analysis, whose accuracy comes from HAL's 
determinism and mode declarations, has been integrated in the HAL compiler
and is shown to produce space improvements as well as speed improvements.
Second, we explain how to modify
the classic PARMA trailing scheme to halve its trailing cost. This
technique is illustrated and evaluated both in the context of dProlog and
HAL. Finally, we explain the modifications needed
by the trailing analysis in order to be combined with our modified PARMA
trailing scheme.  Empirical evidence shows
that the combination is more effective than any of the techniques when
used in isolation.

To appear in Theory and Practice of Logic Programming.
\end{abstract}


\section{Introduction}

The logic programming language Mercury~\cite{somogyi95mercury} is 
considerably faster than 
traditional implementations of 
Prolog due to two main reasons. First,
Mercury requires the programmer to provide type, mode
and determinism declarations whose information is used to generate
efficient target code. And second, variables can only be
ground (i.e., bound to a ground term) or new (i.e.,
first time seen by the compiler and hence unconstrained). Since
neither aliased variables nor partially instantiated structures are 
allowed, Mercury does not need to support full unification; only
assignment, construction, deconstruction and equality testing for
ground terms are required. Furthermore, it does not need to perform trailing,
a technique that allows an execution to resume computation from a
previous program state: information about the old state is logged during
forward computation and used to restore it during backtracking.
This usually means recording the state of unbound variables right before they
become aliased or bound. Since Mercury's new 
variables have no run-time representation
they do not need to be trailed.

HAL~\cite{demoen99overview,flops2002} is a constraint logic language
designed to support the construction, extension and use of constraint
solvers.  HAL also requires type, mode and determinism declarations and
compiles to Mercury so as to leverage from its sophisticated compilation
techniques. However, unlike Mercury, HAL includes a Herbrand constraint
solver which provides full unification. This solver uses Taylor's PARMA
scheme~\cite{taylorthesis,taylor96parma} rather than the standard WAM
representation~\cite{ait-kaci91wam}. This is because, unlike the WAM, the
PARMA representation of ground terms does not contain reference
chains and, hence, it is equivalent to that of Mercury.
Thus, calls to the Herbrand constraint solver can be replaced by calls to
Mercury's more efficient routines whenever ground terms are being
manipulated.

Unfortunately, the increased expressive power of full unification comes at a
cost, which includes the need to perform trailing.  Furthermore, trailing in
the PARMA scheme is more expensive than in the WAM, both in terms of time
and space.  We present here two techniques to counter the trailing penalty
of the PARMA scheme. 
The first is a trailing analysis that detects and
eliminates at compile-time unnecessary trailings and is suitable for any
system based on the classic PARMA trailing scheme. 
Without other supporting information 
such analysis is rather inaccurate, since little is known at
compile-time about the way predicates are used. 
However, when mode and determinism information is available at
compile-time, as in HAL, 
significant accuracy improvements can be obtained. 
The second technique is a
modified PARMA trailing scheme which considerably reduces the required trail
stack size. This technique can be applied to any PARMA-based system and has
been implemented by us in both dProlog~\cite{dProlog} and the Mercury
back-end of the HAL system. Finally, we detail the modifications required by
our trailing analysis in order to be combined with our modified trailing
scheme. The empirical evaluation of each technique indicates that the
combination of the modified trailing scheme with the trailing analysis
results in a significant reduction of trail size at a negligible time cost.

The rest of the paper proceeds as follows. The next section provides a
quick background on trailing, the classic PARMA scheme, and when 
trailing can be avoided. Section~\ref{requirements} summarizes the information
used by our analyzer to improve its accuracy. Section~\ref{analysis} presents
the \texttt{notrail} analysis domain. Section~\ref{body-class} shows how to
analyze HAL's body constructs. Section~\ref{sec:ntdom:opt} shows how to use
the analysis information to avoid trailing. Section~\ref{improved_trailing} presents 
the modified trailing scheme. Section~\ref{analysis-impr} shows the 
changes required by the analysis to deal with this modified scheme. Section~\ref{results} 
presents the results from the experimental evaluation of each technique.
Finally, future work is discussed in
Section~\ref{future}.

\section{Background}\label{background}

We begin by setting some terminology.  
A \emph{bound} variable is a variable that is bound to some nonvariable
term. An \emph{aliased} variable is unbound and 
equated with some other variable. A \emph{free} variable is unbound and
unaliased. We will also refer to a 
\emph{new} variable, which
is a variable in HAL (and Mercury) which has no run-time representation,
since it is yet to be constrained.

In the WAM, an unbound variable is 
represented by a {\em linear} chain. 
If the variable is free the chain has length
one (a cell containing a self-reference). 
When two free variables are
unified, the younger cell is made to point to the older cell (see Section \ref{cond_vs_uncond} for a discussion of relative cell age). 
These two variables are now aliased. A series of
unifications of free variables thus results in a linear chain of references
of which the last one is a self-reference or, in case the variable becomes
instantiated, a bound term. This representation implies that testing whether 
a (source level) variable is bound or unbound, 
requires dereferencing. Such
dereferencing is necessary during each unification 
and it is thus performed quite often.

\begin{figure}[h!]
\begin{centering}
\subfigure[Initially]{{\epsfig{file=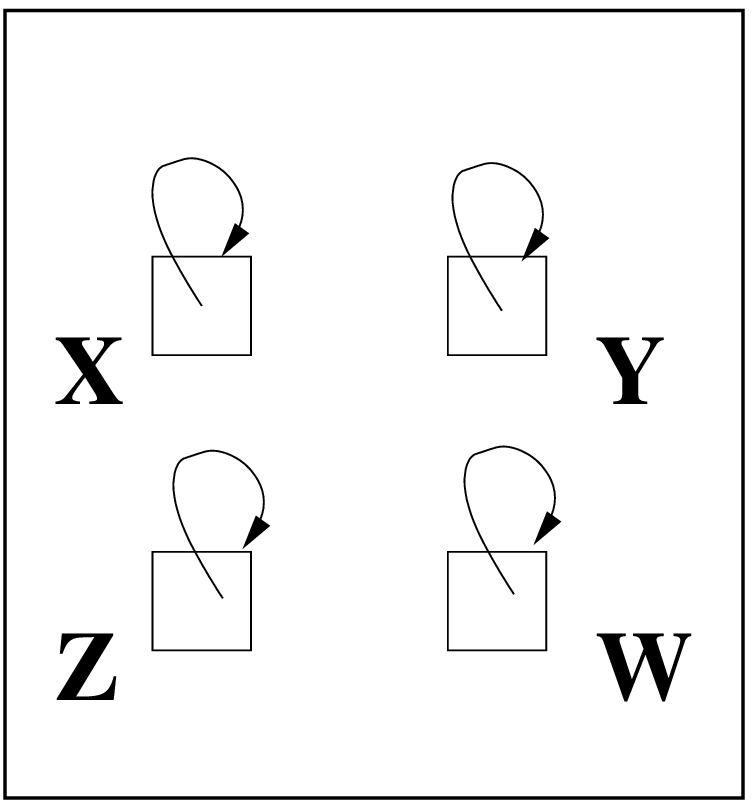,width=.18\textwidth}}}
\subfigure[\texttt{X = Y}]{{\epsfig{file=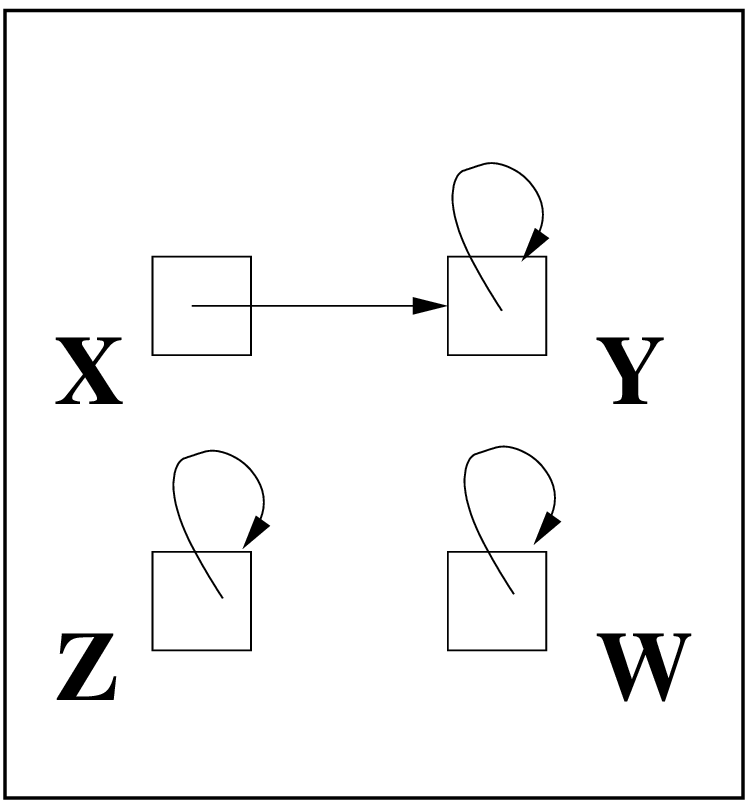,width=.18\textwidth}}}
\subfigure[\texttt{Z = W}]{{\epsfig{file=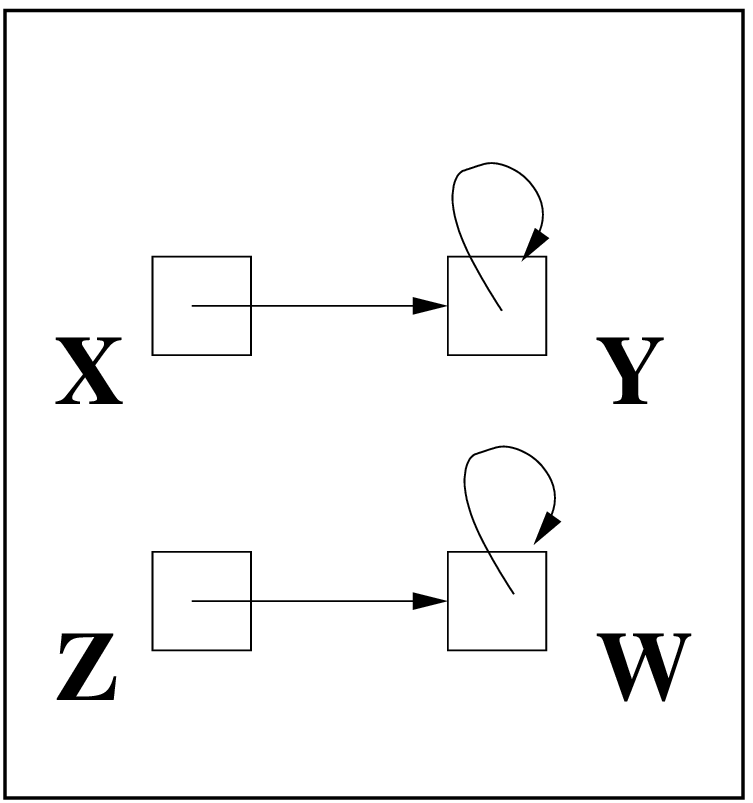,width=.18\textwidth}}}
\subfigure[\texttt{X = Z}]{{\epsfig{file=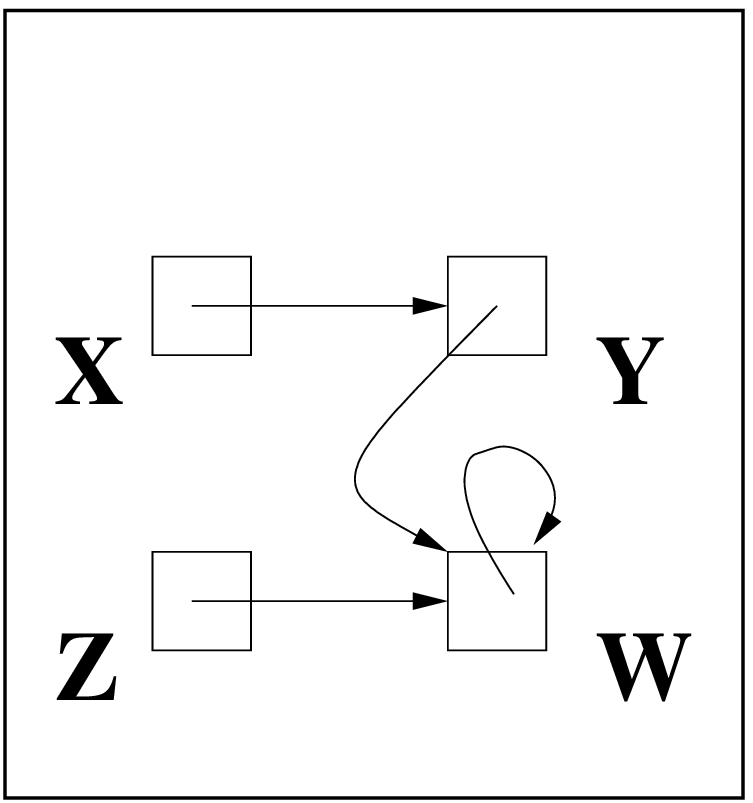,width=.18\textwidth}}}
\subfigure[\texttt{X = a}]{{\epsfig{file=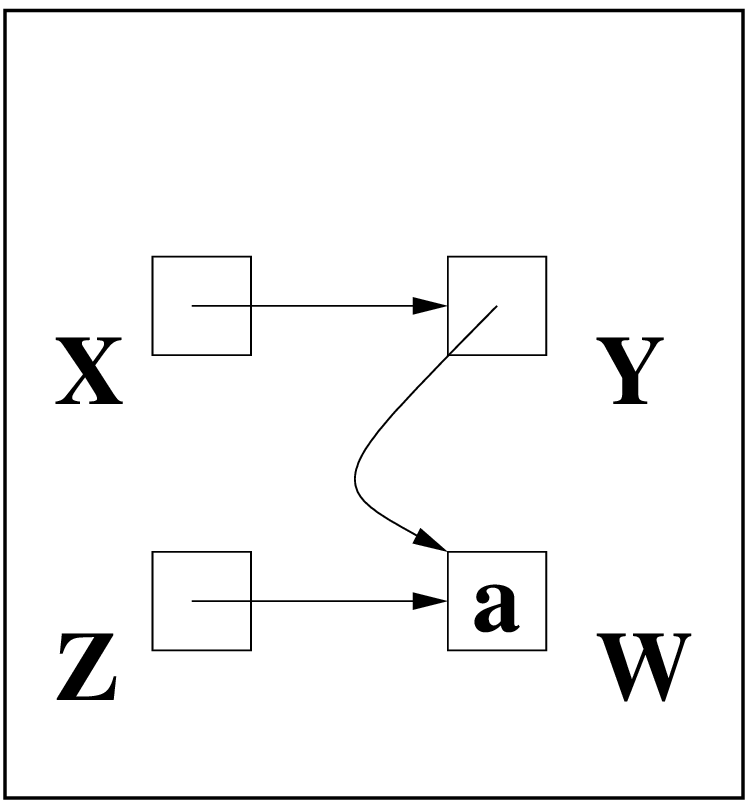,width=.18\textwidth}}}
\caption{Example of binding chains using the WAM representation.}
\label{fig:wam}
\end{centering}
\end{figure}

\begin{example}
Consider the execution of the goal \texttt{X = Y, Z = W, X = Z, X = a} 
when each variable is initially represented by
a self-reference. Using the WAM representation, the first unification points \texttt{X} at \texttt{Y}.
The second unification points \texttt{Z} at \texttt{W}. In the third
unification we must first dereference \texttt{X} to get \texttt{Y},
dereference \texttt{Z} to give \texttt{W}, and 
then point \texttt{Y} at \texttt{W}. In the last unification we
dereference \texttt{X} and set \texttt{W} to \texttt{a}.
The changes in heap states are shown in Figure~\ref{fig:wam}.
\end{example}

In his PARMA-system \cite{taylor96parma}, Taylor 
introduced a different variable
representation scheme that does not suffer from this dereferencing
need. In this scheme an unbound variable is represented by a {\em circular}
chain. If the variable is free the chain has length one (a self-reference
as in the WAM). Unifying two variables in this scheme consists of cutting
their circular chains and combining them into one big circular chain. When
the variable is bound, each cell in the circular chain is replaced by the
value to which it is bound. No dereferencing is required to verify whether
a cell is bound, because the tag in a cell 
immediately identifies the cell as being bound or not. 
However, as we will see later, other costs are incurred by the scheme.

\begin{figure}[h!]
\begin{centering}
\subfigure[Initially]{{\epsfig{file=wam_initial.eps,width=.18\textwidth}}}
\subfigure[\texttt{X = Y}]{{\epsfig{file=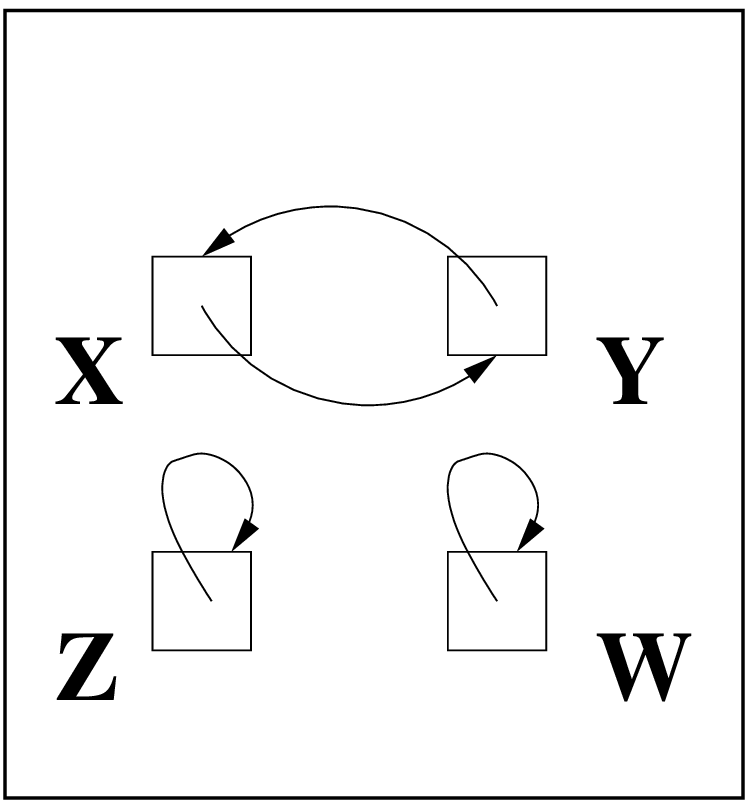,width=.18\textwidth}}}
\subfigure[\texttt{Z = W}]{{\epsfig{file=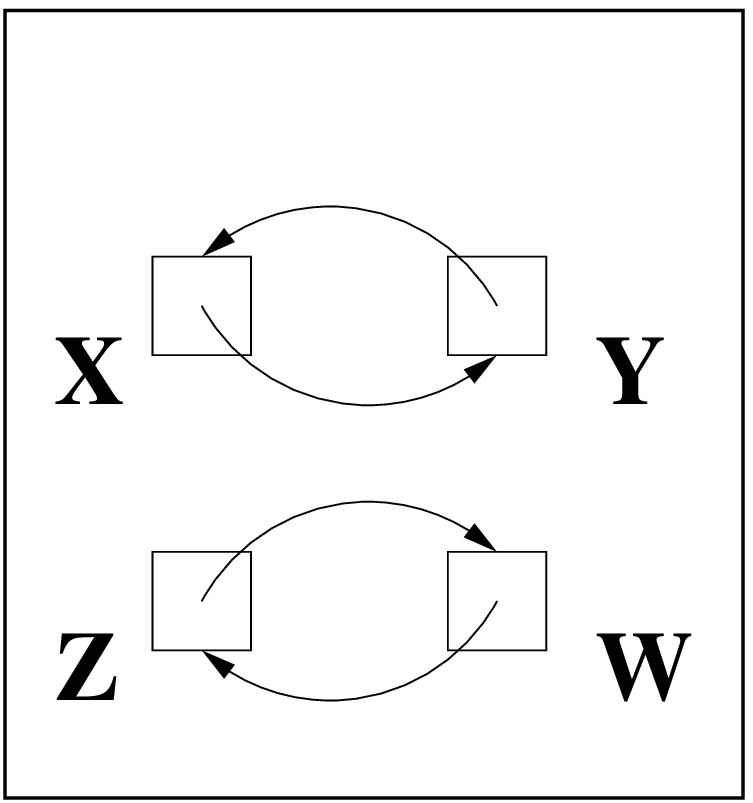,width=.18\textwidth}}}
\subfigure[\texttt{X = Z}]{{\epsfig{file=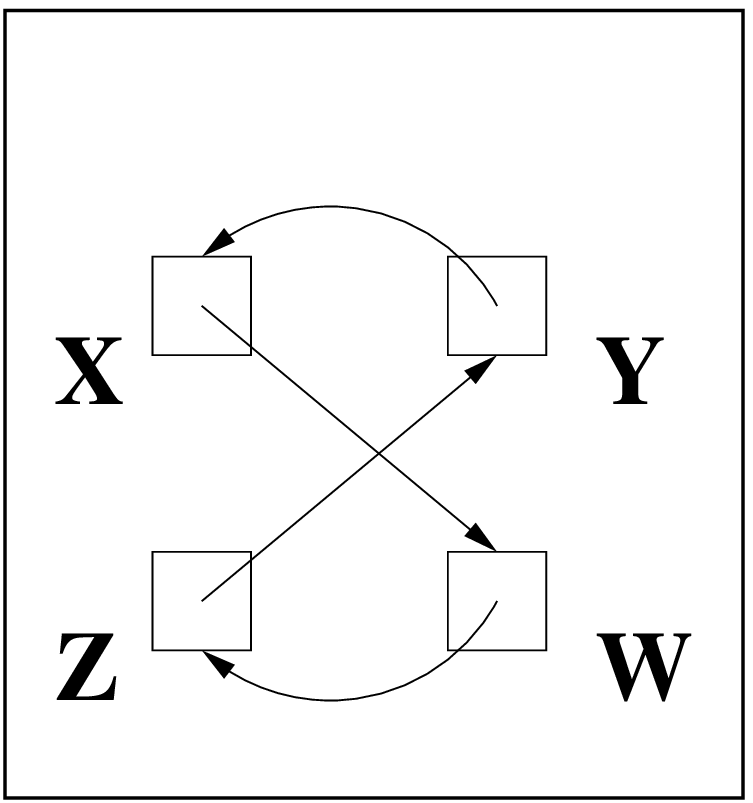,width=.18\textwidth}}}
\subfigure[\texttt{X = a}]{{\epsfig{file=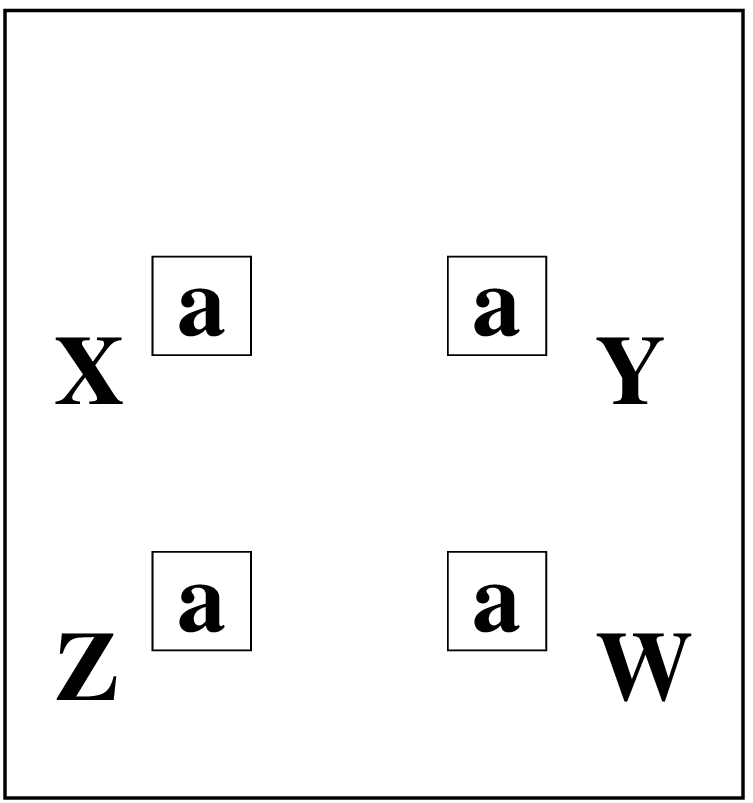,width=.18\textwidth}}}
\caption{Example of binding chains using the PARMA representation.}
\label{fig:parma}
\end{centering}
\end{figure}

\begin{example}\label{ex:parma}
Consider the execution of the same goal 
\texttt{X = Y, Z = W, X = Z, X = a} 
when again each variable is initially represented by
a self-reference. 
Using the PARMA representation, the first unification points \texttt{X} at \texttt{Y}
and \texttt{Y} at \texttt{X}.
The second unification points \texttt{Z} at \texttt{W} and
\texttt{W} at \texttt{Z}. In the third
unification we must point \texttt{X} at \texttt{W}
and \texttt{Z} at \texttt{Y}. In the final unification each
variable in the chain of \texttt{X} is set to \texttt{a}.
The changes in heap states are shown in Figure~\ref{fig:parma}.
Notice how no references remain in the final state, as
opposed to Figure~\ref{fig:wam}(e).
\end{example}

Another difference between the WAM and PARMA binding schemes becomes apparent when
constructing a new term containing an unbound variable $X$.
Effectively, we are aliasing a new variable with $X$
and, hence, this new variable must be added into the variable chain of $X$.

\begin{figure}[h!]
\begin{centering}
\subfigure[\texttt{X = Y}]{{\epsfig{file=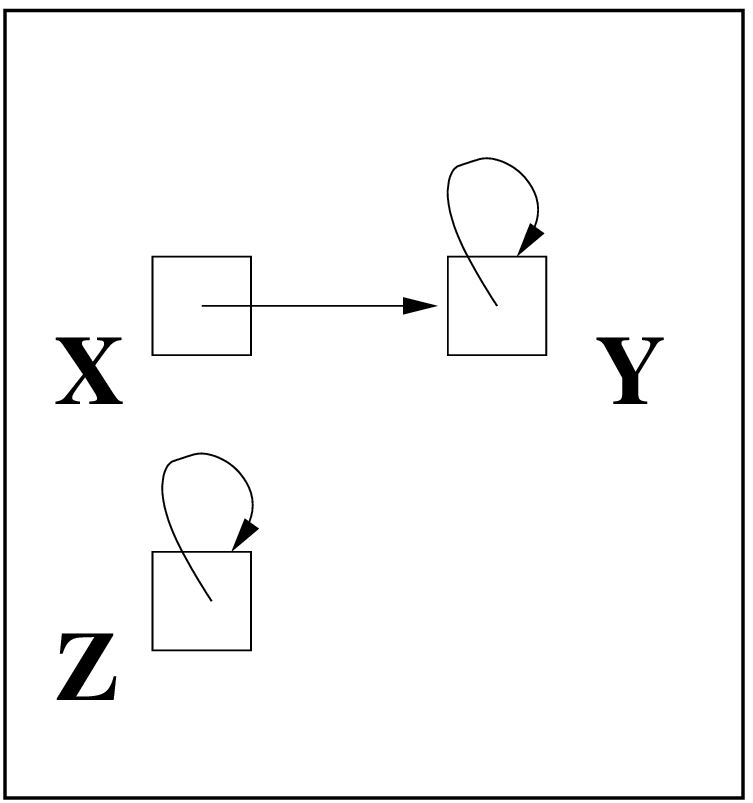,width=.18\textwidth}}}
\subfigure[\texttt{Z=f(X)}]{{\epsfig{file=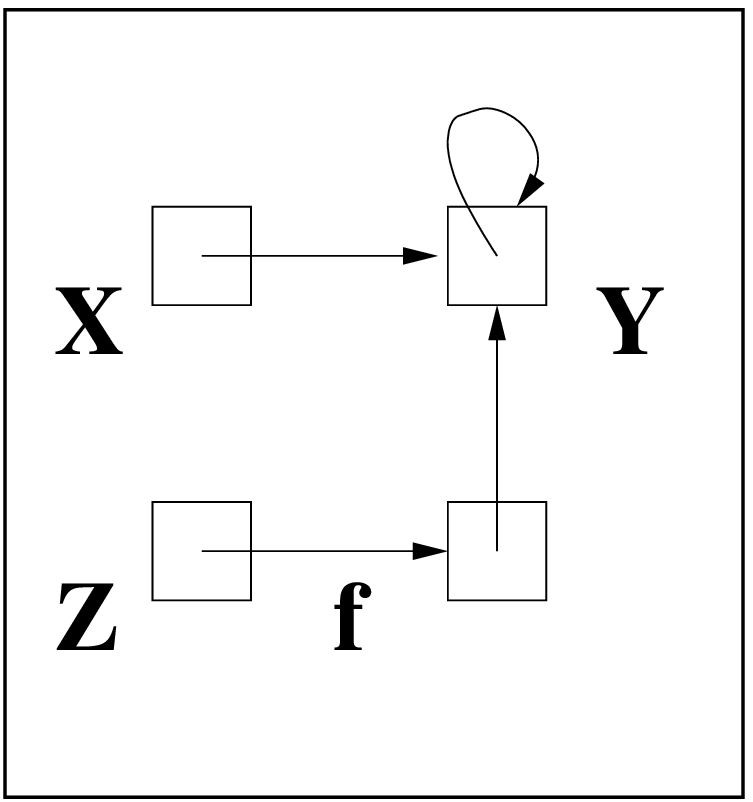,width=.18\textwidth}}}
~~~~~~~~
\subfigure[\texttt{X = Y}]{{\epsfig{file=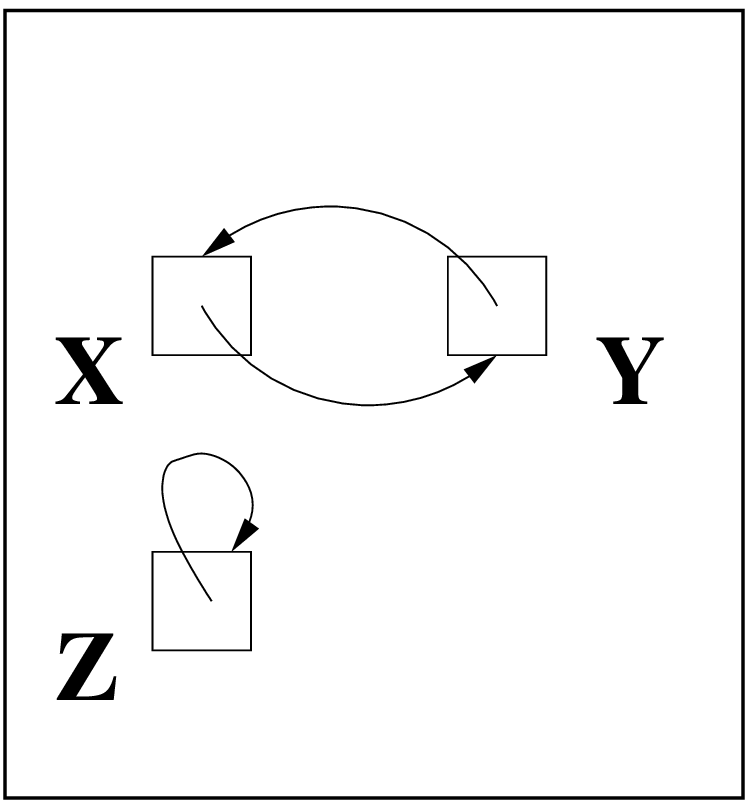,width=.18\textwidth}}}
\subfigure[\texttt{Z=f(X)}]{{\epsfig{file=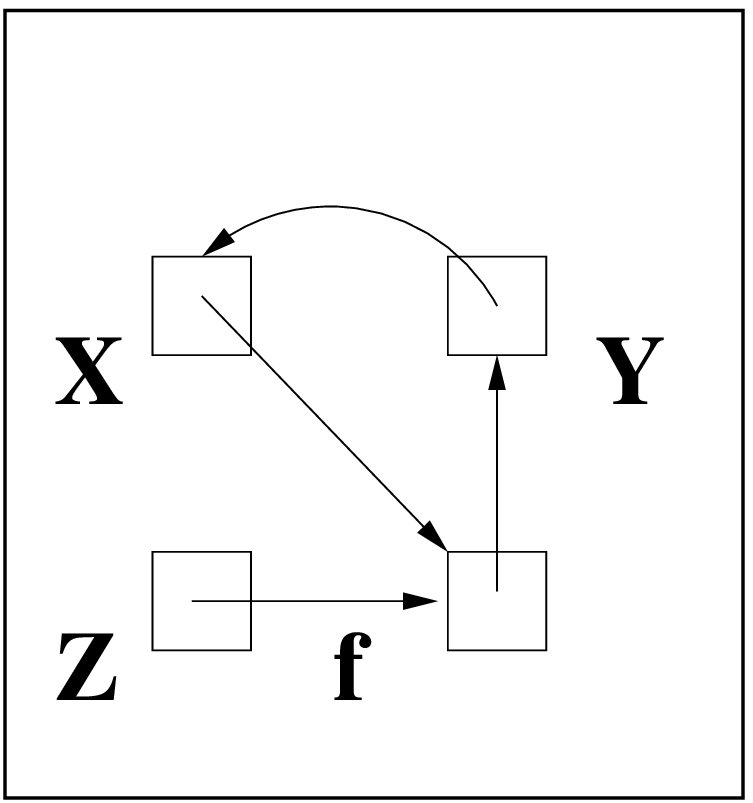,width=.18\textwidth}}}
\caption{Example of constructing a term containing an unbound variable
using both WAM (a)(b) and PARMA (c)(d) representations.}
\label{fig:unbound}
\end{centering}
\end{figure}

\begin{example}\label{ex:unbound}
Consider the execution of the goal \texttt{X = Y, Z = f(X)} when each variable
is initially represented by a self-reference.

Using the WAM representation, the first unification points \texttt{X} at
\texttt{Y}. The second unification constructs a heap term \texttt{f(X)} with
the content of \texttt{X}, namely \texttt{Y}, and points \texttt{Z} at this.

Using the PARMA representation, the first unification chains \texttt{X} and
\texttt{Y} together. The second unification has to add the copy of \texttt{X}
in \texttt{f(X)}, to the chain for \texttt{X}. The resulting heap states
are shown in Figure~\ref{fig:unbound}.
\end{example}

As mentioned before, trailing is a technique that stores enough 
information regarding
the representation state of a variable before each choice-point, to be able
to reconstruct such state upon backtracking. For both WAM and PARMA chains
the change of representation state occurs at the cell level: from being a
self-reference (when the variable represented by the cell -- {\em the
associated variable} -- is unbound and unaliased), to pointing to another
cell in the chain (when the associated variable gets aliased), to pointing
to the final bound structure (when the variable is bound directly or 
indirectly). Thus, what needs to be trailed are the cells.

In the rest of the section we will discuss the PARMA trailing scheme in
greater detail, the orthogonal issue of 
conditional/unconditional trailing,
and a possible improvement based on 
compile-time detection of unnecessary trailings. 

\subsection{The classic PARMA Scheme: Value trailing}

The classic PARMA trailing scheme uses \emph{value trailing}, described by
the following C-like code:\footnote{All code in this paper is 
pseudo-C code. Implementation details that obfuscate rather than clarify the
concepts at hand, have been omitted.}
\begin{Verbatim}
         valuetrail(p) { 
               *(tr++) = *p; /* store the contents of the cell p */   
               *(tr++) = p;  /* store the address of the cell p  */
         }        
\end{Verbatim}
which takes the address \texttt{p} of a cell in a PARMA chain and stores in
the trail stack first the (old) contents of the cell and then its address.
Here, \texttt{tr} is a global pointer to the top of the trail stack. 

The untrail operation for value trailing is straightforwardly defined by:
\begin{Verbatim}
         untrail_valuetrail() {
               address = *(--tr);  /* retrieve the cell address */
               *address = *(--tr); /* recover the cell contents */
         }
\end{Verbatim}
which first pops the address of a cell and then its contents.

In contrast, trailing in the WAM stores only the address of the cell.  The
reasons are twofold. First, a cell is updated at most once, from a
self-reference to a pointer to either another cell in a linear chain or a
structure. And second, for a self-referencing cell the address and the
content of the cell are the same. Therefore, when a cell is updated the old
content of the cell (which is the one stored during trailing) is always the
same as its address. This allows the WAM value trailing to be optimized by
only storing the address of the cell, reducing by half the space cost of a
single cell trailing.

Let us now discuss when cells need to be trailed in the classic PARMA
scheme. We have seen before that trailing is only needed when the
representation state of a variable changes, and that this can only happen
when the variable is unbound and, due to a unification, it becomes either
aliased or bound. Therefore, we only need to trail cells when their
associated variables are involved in a unification or when creating a new term which contains an unbound variable.
The following discussion distinguishes three cases: cells associated to variables
involved in a variable--variable unification, in a
variable--nonvariable unification, and in new term construction.

        \paragraph{Trailing during variable--variable unification:}

        The result of aliasing two unbound variables belonging to separate
chains is the merging of the two chains into a single one. This can be
done by changing the state of only two cells: those associated to each of
the variables. Since each associated cell appears in a different chain, the
final chain can be formed by simply interchanging their respective
successors. One can then reconstruct the previous situation by remembering
which two cells have been changed and what their initial value was. This is
achieved for unification {\tt X = Y} by the following (simplified) code:
\begin{Verbatim}[xleftmargin=30mm]
valuetrail(X);
valuetrail(Y);
tmp = *X;
*X = *Y;
*Y = tmp;
\end{Verbatim}

        Notice that \texttt{X} and \texttt{Y} are trailed independently.
As only their associated cells need to be trailed, we will refer to this
kind of trailing as {\em shallow} trailing.

In contrast, for this kind of unification the WAM will update and trail the last cell 
in just one of the two linear chains. Hence, the space cost is four times lower (one value as opposed to four).

\begin{example}\label{ex:basictrail}
Consider the PARMA trailing that occurs during the first three
unifications of the goal
\texttt{X = Y, Z = W, X = Z, X = a} from Example~\ref{ex:parma}, when each variable is initially represented by
a self-reference.
From the first unification 
we trail \texttt{X} together with its initial value (which, since \texttt{X} is a self-reference, is also) \texttt{X},
and \texttt{Y} together with its initial value \texttt{Y}.
Similarly, for the second unification we trail 
\texttt{Z} together with its value \texttt{Z}, and \texttt{W} together with its value \texttt{W}.
For the third unification, we trail \texttt{X} together with its value \texttt{Y}, and 
\texttt{Z} together with its value \texttt{W}.
The resulting trail is
\begin{center}
\fbox{\texttt{X}} \fbox{\texttt{X}} \fbox{\texttt{Y}} \fbox{\texttt{Y}} 
\fbox{\texttt{Z}} \fbox{\texttt{Z}} \fbox{\texttt{W}} \fbox{\texttt{W}} 
\fbox{\texttt{Y}} \fbox{\texttt{X}} \fbox{\texttt{W}} \fbox{\texttt{Z}} 
\end{center}
The WAM trail for the same goal illustrated in
Figure~\ref{fig:wam} trails first \texttt{X}, then
\texttt{Z} and finally \texttt{Y}. 
The resulting trail is \fbox{\texttt{X}} \fbox{\texttt{Z}} \fbox{\texttt{Y}}.
\end{example}

        \paragraph{Trailing during variable--nonvariable unification:}

        When an unbound variable becomes bound, every single cell in its
chain is set to point to the nonvariable term. Thus, we can only
reconstruct the chain if {\em all} cells in the chain are trailed. The
combined unification-trailing (simplified) code for unification 
\texttt{X = Term} is
as follows:
\begin{Verbatim}[xleftmargin=30mm]
start = X;
do {
   next = *X;
   valuetrail(X);
   *X = Term;
   X = next;
} while (X != start);
\end{Verbatim}

Since all cells in the chain of the unbound variable are trailed, we will
refer to this kind of trailing as {\em deep} trailing. 

In contrast, for this kind of unification, the WAM will trail again just 
one cell in the
linear chain.  Hence, the space complexity for WAM is just $\mathcal{O}(1)$
compared to $\mathcal{O}(n)$ for PARMA, where $n$ is the number of cells
in the chain. However, the time complexity is $\mathcal{O}(n)$ for both,
due to the dereferencing in the WAM.

\begin{example}\label{ex:more}
Consider the PARMA trailing that happens in the last unification
\texttt{X = a} of the goal from Example~\ref{ex:parma}.
The binding of all variables in the chain 
adds the trail elements
\begin{center}
\fbox{\texttt{W}} \fbox{\texttt{X}} \fbox{\texttt{X}} \fbox{\texttt{Y}} 
\fbox{\texttt{Y}} \fbox{\texttt{Z}} \fbox{\texttt{Z}} \fbox{\texttt{W}} 
\end{center}
In contrast the WAM trailing adds a single trail element 
\fbox{\texttt{Y}}.
\end{example}

        \paragraph{Trailing during new term construction:}

	As mentioned before, when a new term is constructed on the heap with a copy of an
unbound variable $X$, the cell containing this copy must be added into
the chain for $X$.  This means we must trail
$X$ since its value (i.e., its successor in the chain) is going to change.  We do not need to trail the new cell since
it clearly has no previous value we need to recover.
The combined construction-trailing (simplified) code for
constructing \texttt{f(X)} where \texttt{X} is an unbound variable 
and \texttt{th} is the current top
of heap pointer, is:
\begin{Verbatim}[xleftmargin=30mm]
*(++th) = *X;
valuetrail(X);
*X = th;
\end{Verbatim}
In contrast, for this construction the WAM need not trail
at all since it simply points the new cell at the old unbound variable.

If \texttt{X} is either a bound or a new variable, this complexity does not
arise: \texttt{X} will be placed in the new structure pointing to either
the nonvariable term or to itself, with no trailing required in any case.
 
        \paragraph{Summary:}

The major advantage of the PARMA binding scheme is that it
requires no dereferencing, while its major disadvantages are (for
a detailed account see \cite{lindgren95taylor}):
\begin{enumerate}
	\item PARMA trails more cells per unification: two in variable-variable unifications and all in variable-nonvariable, versus one.
	\item Trailing of an individual cell is more expensive: two slots used versus one.
	\item Unlike in the WAM, cells can be trailed more than once: every time a cell is updated which can happen more than once.
	\item Copying an unbound variable into a structure involves trailing a cell. 
\end{enumerate}
As a result, the trail stack usage is expected to be much higher in the
PARMA scheme than in the WAM. Demoen and Nguyen \cite{dProlog} have indeed
observed in the dProlog system maximal trail sizes for the PARMA scheme
that are on average twice as large as with the WAM scheme. The techniques
we present in this paper attempt to counter the disadvantages. The trailing
analysis reduces the number of trailings and thereby counters disadvantages
1, 3 and~4, while the modified trailing scheme counters disadvantage 2.

\subsection{Conditional versus unconditional trailing}\label{cond_vs_uncond}

A cell that is changed only requires trailing if the cell did exist before
the most recent choice point since, otherwise, there is no previous state
that has to be restored during backtracking. This property applies equally
to the WAM and PARMA schemes.

In some systems a simple run-time test can be used to verify whether a cell
is older than the most recent choice point. Younger cells
require no trailing. 
If all cells on the heap are kept in order of allocation, the
test simply checks whether the address of the cell is smaller than that of
\texttt{bh}, the address of the top of the heap at the beginning of the
most recent choice point. 
Systems, such as dProlog, which take advantage of
this property use what is known as {\em conditional trailing}.
Let us assume the existence of function \texttt{is\_older(p,bh)}
which succeeds if \texttt{p < bh}. Conditional trailing is then described by
the following code:
\begin{Verbatim}
         cond_valuetrail(p, bh) { 
            if (is_older(p,bh)) 
                valuetrail(p);
         }
\end{Verbatim}
thus avoiding the trailing of cells which are newer than the most recent
choice point. The code for variable--variable and variable--nonvariable
unification described in the previous sections using the unconditional {\tt valuetrail}
operation can be rewritten to use conditional trailing by simply
substituting each call to {\tt valuetrail} by a call to {\tt
cond\_valuetrail}. The untrail operation remains unchanged. 

In systems where the order of cells on the heap is not guaranteed,
unconditional trailing is required. The Mercury back-end of the HAL system, for example, is
such a system since Mercury uses the Boehm garbage collector which does not
preserve the order of the cells on the heap between garbage collections.
Other systems use unconditional trailing at least during some unifications
(see for instance \cite{Aquarius}). In \cite{dProlog} it is shown that
global performance is hardly affected by the choice between conditional or
unconditional trailing, since the savings made on avoided trailings are balanced
by the overhead of the run-time tests.  

The differences between conditional and unconditional trailing do not
affect the proposed analysis. Thus, the same analysis can still be used if
at some point conditional trailing becomes 
available in Mercury.

\subsection{Unnecessary trailing in the classic PARMA scheme:} 

When considering 
the trailing of an unbound variable 
appearing in a unification, there are at least
two cases in which its trailing can be avoided:

\begin{itemize}
\item 
If the variable is new 
there is no previous value to remember and, therefore,
trailing is not required. 
This is in fact a subset of the cases exploited by
conditional trailing.

\item The cells that need to be trailed (the associated cell  in the case
of variable--variable, all cells in the case of variable--nonvariable)
have already been trailed {\em since the most recent choice-point}.
Upon backtracking only the earliest trailing after the choice-point is
important, since that is the one which enables the reconstruction of the
state of the variable before the choice-point.
\end{itemize}

In the following sections we will see how compile-time analysis information
can be obtained to detect the above two cases and can therefore be used to
(a) eliminate unnecessary trailing in the classical PARMA trailing scheme,
and (b) eliminate run-time tests performed by conditional trailing on
variables known at compile-time to have no representation and thus be
younger than the most recent choice point.

\section{Language Requirements}\label{requirements}

The analysis presented in this paper was designed for the HAL
language. However, it can be useful for any language that uses PARMA
representation and that provides accurate information regarding the
following properties:

\begin{itemize}
\item Instantiation state: trailing analysis can gain accuracy by taking into
account the instantiation state of a program variable, 
i.e. whether the variable is new, ground or old. 
State new corresponds to program 
variables with no internal representation
(equivalent to Mercury's free instantiation). State ground corresponds to
program variables 
known to be bound to ground terms. In any other case the state
is old, corresponding to program 
variables which might be unbound but do have a
representation (a chain of length one or more) or bound to a term not known
to be ground. Program 
variables with instantiation state new, ground or old will
be called new, ground or old variables, respectively.  Note that once a new
variable becomes old or ground, it can never become new again. And once it
is known to be ground, it remains ground. Thus, the three states
can be considered mutually exclusive. 
The information should be available at
each program point $p$ as a table associating 
with each variable in scope of $p$
its instantiation state.

We will represent the instantiation table
information at program point $p$ as follows.
Let $Var_p$ denote the set of all program variables in scope at 
program point $p$.
The function $inst_p : Var_p \rightarrow \{new,ground,old\}$
defines the instantiation state of program variable $X$ at point $p$.
This function allows us to partition
$Var_p$ into three disjoint sets: $New_p$, $Ground_p$ and $Old_p$ containing
the set of new, ground and old variables, respectively. 

\item Determinism: trailing analysis can also gain accuracy from the
knowledge that particular predicates have at most one solution.  This
information should be available as a table associating with each predicate
(procedure to be more precise) its determinism. Herein
we will refer to six main kinds of determinism: \texttt{semidet}
(minimum-maximum set of solutions: 0-1), \texttt{det} (1-1), \texttt{multi}
(1-$\infty$), \texttt{nondet} (0-$\infty$), \texttt{erroneous} (1,0),
and \texttt{failure} (0-0).

For our purposes we will only be interested in whether a predicate
can return more than one answer.
We will represent the determinism table by a function
$det : Pred \rightarrow \{0,1,\infty\}$ 
which maps each predicate $q$ to its maximum number of solutions.

\item Sharing: trailing analysis can exploit sharing information to
increase accuracy. This information should be available at each program
point $p$ as a table associating with each variable in scope of $p$ the set of
variables which possibly share with it.
Clearly, any variables that may be aliased together must possibly share.

We will represent the sharing table at program point $p$
by the function
$share_p : Old_p \rightarrow {\cal P}(Old_p)$ 
which assigns to each program variable in $Old_p$ the set
of program variables in $Old_p$ that share with it. Note that
program variables in $New_p$ and $Ground_p$ cannot share by definition.
\end{itemize}

\section{The \texttt{notrail} Analysis Domain}\label{analysis}

The aim of the \texttt{notrail} domain is to keep enough information to be
able to decide whether the {\em run-time variables} in a unification need to be
trailed or not, so that if possible, optimized versions which do not perform
the trailing can be used instead. In order to do this, we must remember that
only run-time variables which are unbound and have a representation (i.e., are not new) need to be trailed.  
This suggests
making use of the instantiation information mentioned in the previous section.
Note that, since the analysis works on the level of \emph{program variables},
some indirection will be required.

We have already established that program variables in $New_p$ and $Ground_p$
represent run-time variables which do not need to be trailed. Thus, only
variables in $Old_p$ need to be represented in the \texttt{notrail} domain.
the set of new, ground and old program variables, respectively.  Assuming that
$Var_p$ contains $n$ variables and the tree we have used to implement the
underlying table is sufficiently balanced, then, the size of the $Old_p$
is $\mathcal{O}(n)$ and the complexity of $inst_p$ is $\mathcal{O}(\log{n})$.

Recall that $Old_p$ contains all program variables representing not only 
run-time variables which are unbound and have a representation, but also run-time variables 
bound to terms which the analysis cannot ensure to be
ground. This is necessary to ensure correctness: even though run-time variables which
are bound do not need to be trailed, the
nonvariable terms to which they are bound might contain one or more unbound run-time variables. It is
the trailing state of these unbound run-time variables that is represented
through the domain representation of the bound program variable.

Now that we have decided which program variables need to be represented by
our domain, we have to decide how to represent them. We saw before that it
is unnecessary to trail a run-time variable in a variable--variable unification if
its associated cell has already been trailed, i.e., if the run-time variable has
already been shallow trailed since the most recent choice-point. For the
case of variable--nonvariable unification this is not enough, we need to
ensure all cells in the chain have already been trailed, i.e, the run-time variable
has already been deep trailed. This suggests a domain which distinguishes
between shallow and deep trailed run-time variables. This can be easily done by
partitioning $Old_p$ into three disjoint sets of program variables with a different
trailing state: those representing run-time variables which might not have been trailed yet, those representing run-time variables which
have at least been shallow trailed, and those representing run-time variables which have been deep
trailed. It is sufficient to keep track of only two sets to be able to
reconstruct the third. Hence, the type of the elements of our
\texttt{notrail} domain \ntdom{} will be $\mathcal{P}(Old_p) \times
\mathcal{P}(Old_p)$, where the first component contains the set of
program variables representing run-time variables which have already been shallow trailed, and the second component
contains the set of program variables representing run-time variables which have already been deep trailed. In the following we
will use $l_1,l_2,\ldots$ to denote elements of \ntdom{} at program points
$1, 2, \ldots$, and $s_1, s_2, \ldots$ and $d_1,d_2,\ldots$ for the
already shallow and deep trailed components of the corresponding
elements. Also, the elements of the domain will be referred to as
descriptions, with descriptions before and after a goal being referred to as
the pre- and post-descriptions, respectively.

Note that, by definition, we can state that if a run-time variable has already been
deep trailed, then it has also been shallow trailed (i.e., if all cells in the
chain have already been trailed, then the cell associated to the variable
has also been trailed).  The partial ordering relation $\sqsubseteq$ on
\ntdom{} is thus defined as follows:
\begin{multline*}
\forall (s_p^1,d_p^1),(s_p^2,d_p^2) \in L_{notrail} : 
(s_p^1,d_p^1) \sqsubseteq (s_p^2,d_p^2) \Leftrightarrow \left\{\begin{array}{lclllll} s_p^2  & \subseteq & d_p^1 & \cup & s_p^1 \\
d_p^2 & \subseteq & d_p^1  \end{array}\right.
\end{multline*}

\begin{figure}
        \input{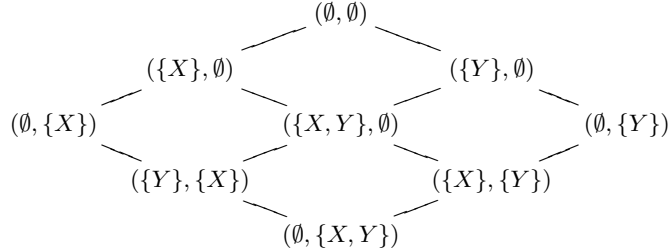}
        \caption{\texttt{Notrail} lattice Hasse diagram
for variables $\{X,Y\}$ where if $l_1 \sqsubseteq l_2$ then
$l_1$ is below $l_2$ in the diagram }\label{lattice}
\end{figure}

This implies that deep trailing is stronger information than shallow
trailing, and shallow trailing is stronger than no trailing at all. Also
note that descriptions are compared at the same program point only (so that
the instantiation and sharing information is identical).  
An example of a trailing
lattice is shown in \figurename{}~\ref{lattice}.  Clearly $(L_{notrail},
\sqsubseteq)$ is a complete lattice with top description
$\top_p = (\emptyset,\emptyset)$ and bottom description $\bot_p =  (\emptyset, Old_p)$.

There are two important points that need to be taken into account when
considering the above domain. 
The first point is that the $d_p$ component
of a description will be used not only to represent already deep trailed
variables but any variable in $Old_p$ which, for whatever reason (e.g., it has been initialized since the last choicepoint), does not
need to have any part of it trailed

The second point is that as soon as a 
deeply trailed program  variable $X$ is made to
share with a shallow trailed program 
variable $Y$, $X$ also must become shallow
trailed since some cell in some newly merged chain might come from $Y$ and
thus might not have been trailed. 
The sharing information at each program point is 
used to define the following function which
makes trailing information consistent with its associated sharing
information:
$$consist_p((s,d)) = (s \cup x, d \setminus x )$$
where
\[x = \{ X \in d | (share_p(X) \setminus d) \not = \emptyset \}\]
Intuitively, the function eliminates from $d$ every program variable $X$
which shares with other variables not in $d$, and adds them to $s$.
From now on we will assume that $\forall (s,d) \in \ntdom{}:
consist_p((s,d)) = (s,d)$ and use the $consist$ function to preserve
this property.\footnote{Note that the \texttt{notrail} domain can  be
seen as a ``product domain'' that also includes the mode and sharing
information. However, for simplicity, we will consider the different
elements separately, relating them only via their associated program
point.}

Given HAL's implementation of the sharing analysis domain
\texttt{ASub}~\cite{sonder86} the time complexity of 
determining $share_p(X)$ for a variable $X$ is $\mathcal{O}(n^{2})$.
Furthermore, since \texttt{ASub}
explicitly carries the set of ground variables at each program
point ($g_p$), we will use 
this set rather than computing a new one ($Ground_p$)
from the instantiation information, thus increasing
efficiency. The major cost of
$consist_p$ is the computation of $x$: for each of the $\mathcal{O}(n)$
variables the $share_p$ set has to be computed. All other set operations
are negligible in comparison. Hence, the overall time complexity is
$\mathcal{O}(n^{3})$.  We will see that the complexity of this function
determines the complexity of all the operations that use it. Thus, we
will use it only when strictly necessary.

In summary, each element $l_p=(s_p,d_p)$ in our domain can be interpreted
as follows. Consider a program variable $X$. 
If $X \in d_p$, this means 
that all cells in all chains represented 
by $X$ have already been trailed
(if needed). 
Therefore, $X$ does not need to be trailed in any unification
for which $l_p$ is a pre-description.  
Note that $X$ could be a bound variable which includes
many different variable chains.
If $X \in s_p$ we have two
possibilities. 
If $X$ is known to be unbound, then its associated cell has
been shallow trailed. Therefore, it does not need to be trailed in any
unification for which $l_p$ is a pre-description (although, in practice, we will only
consider optimizing variable-variable unifications). 
If $X$ might be bound, then a cell of one of its chains might not be
trailed. As a result, no optimization can be performed in this case.

We could, of course, represent bound variables more accurately, by
requiring the domain to keep track of the different chains contained in the
structures to which the program variables are bound, their individual
trailing state and how these are affected by the different program
constructs.  Known techniques 
(see for instance \cite{gerda,pascal,mulkers,lagoontypeframe}) 
based on type information could be used to keep track of the
constructor that a variable is bound to and of the trailing state of the
different arguments, thereby making this approach possible.

        \section{Analyzing HAL Body Constructs with \ntdom{}}\label{body-class}

This section defines the \texttt{notrail} operations required by HAL's
analysis framework~\cite{modular-anal-lopstr-formal,nethercote01thesis} 
to analyze the different body
constructs. This framework is quite similar to the well known framework of
\cite{bruy91} when analyzing a single module. 
While the analysis framework handles analysis of multiple module programs,
it makes no extra demands on the analysis domain. Thus, for this paper we will simply treat
the program to be analyzed as a single module.
For each body construct in HAL, we will show how to obtain the
post-description from the information contained in the pre-description.

                \paragraph{Variable initialization} $init(X)$

In HAL a variable $X$ transits from its initial instantiation
new to instantiation old by being initialized.  
Since a new variable does not need to be trailed, we can simply add $X$ to the $d$ component of the pre-description (recall that $d$ not only represents already deep trailed variables, but also any other old variable which does not need to be trailed). Formally, let $l_1=(s_1,d_1)$ be the pre-description,
the post-description $l_2$ can be obtained as:
\[
l_2 = (s_1, d_1 \cup \{X\})
\]

        \paragraph{Variable--variable unification: $X = Y$.}

There are several cases to consider:
\begin{itemize}
\setlength{\itemsep}{0mm}
\setlength{\parskip}{0mm}
\item
If one of the variables (say $X$) is new, it will simply be
assigned a copy of the pointer of $Y$. After the unification is 
performed, the trailing state of $X$ becomes that of $Y$. Thus, the trailing state of
$X$ in the post-description should be that of $Y$ in the pre-description. 
Note that this will never require a
call to $consist$ since a new variable cannot
introduce any sharing.

\item
If one of the variables is ground, the other one will be
ground after the unification. Hence, neither of them will appear
in the post-description.

\item
If both variables are deep trailed, all cells in their associated
chains are trailed and will remain trailed after
unification (which is obtained by simply merging the chains).
Hence, all variables retain their current trailing state and the pre-description will remain unchanged.

\item
If both variables are already aliased (they belong to the same chain) nothing is done by
unification. Hence, they will retain the current trailing state.
Hence, all variables retain their current trailing state and the pre-description will remain unchanged.

\item
Otherwise, at least one of the variables is not deep trailed and two unaliased variables are being considered.
If both variables are unbound, 
unification will merge both chains while at the same
time performing shallow trailing if necessary.
Thus, after the unification
both variables will be shallow trailed.  
If at least one variable
is bound, the other one will become bound after the unification. 
As stated earlier, bound variables can be treated in the same way.

Note that if either variable was deep trailed before the unification,
all shared variables must become shallow trailed as well after the
unification. This requires applying the $consist$ function.
\end{itemize}

Formally, let $l_1=(s_1,d_1)$ be the pre-description  and $g_2$ be the
set of ground variables at program point $2$ after the
unification. Its post-description $l_2$ can be obtained as:
\[ l_2 = \mathit{unify}(X,Y) = \left\{\begin{array}{l@{~~}l}
same(X,Y,l_1)		& X \mbox{\ is\ new}		\\
remove\_ground(l_1,g_2) & X \mbox{\ is\ ground}		\\
min(X,Y,l_1)		& X \mbox{\ and\ } Y \mbox{\ are\ old}	\\
\mathit{unify}(Y,X)	& \mbox{otherwise} 		\\
\end{array}\right.\]
with
\[\begin{array}{rcl}

same(X,Y,(s_1,d_1))&\hspace{-3mm}=\hspace{-3mm}& \left\{\begin{array}{l@{~~}l} (s_1 \cup
\{X\},d_1)&  Y \in s_1 \\
(s_1,d_1 \cup \{X\})& Y \in d_1 \\
(s_1,d_1)& \mbox{otherwise} \\
\end{array}\right.\\

remove\_ground(l_i,v_i) &\hspace{-3mm}=\hspace{-3mm}&(s_i \setminus v_i, d_i \setminus v_i) \\

min(X,Y,(s_1,d_1)) &\hspace{-3mm}=\hspace{-3mm}& \left\{\begin{array}{l@{~~}l}
(s_1,d_1) & \{X,Y\} \subseteq d_1 \\
consist_2((s_1 \cup \{X,Y\}, d_1 \setminus \{X,Y\})) & X \not\in share_1(Y)  \\ (s_1, d_1)     & \mbox{otherwise}  \\ 
\end{array}\right. 
\end{array}\]
Here $same(X,Y,l_i)$ gives $X$ the same trailing state as $Y$,
$remove\_ground(l_i,v_i)$ removes all variables in $v_i$ from $l_i$,
and $min(X,Y,l_i)$ distinguished between three cases. If $X$ and $Y$ are both deep trailed, nothing has to be changed.
If $X$ and $Y$ are definitely not aliased (they do not share)
it ensures that they move to a shallow trailed state.
Otherwise, the description must remain unchanged since unification might have done nothing (and thus
they might still be untrailed, so adding them to $s_1$ would be a mistake). Note that there is no need to
apply $consist$ here since $X$ and $Y$ already share in the pre-description and, although sharing information might have changed,
it can only create sharing among variables already connected (through $X$ and $Y$) by the closure under union performed by $consist$.

The worst case time complexity, 
$\mathcal{O}(n^{3})$, is again due to $consist$.

\paragraph{Variable--term unification: $Y = f(X_1,\ldots,X_n)$.}

There are two cases to consider: 
If $Y$ is new, the unification simply constructs the term in
$Y$. 
Otherwise, we can treat this for the purpose of the analysis as two unifications,
$Y' = f(X_1,\ldots,X_n), Y = Y'$ where $Y'$ is a new variable.
Since unifications of
the form $Y' = Y$ have been discussed above, here we only focus on the
construction into a new variable. In the following we assume that the $Y$
in the variable-term unification is new.

\begin{figure}
\centerline{
\subfigure[Before.]{{\epsfig{file=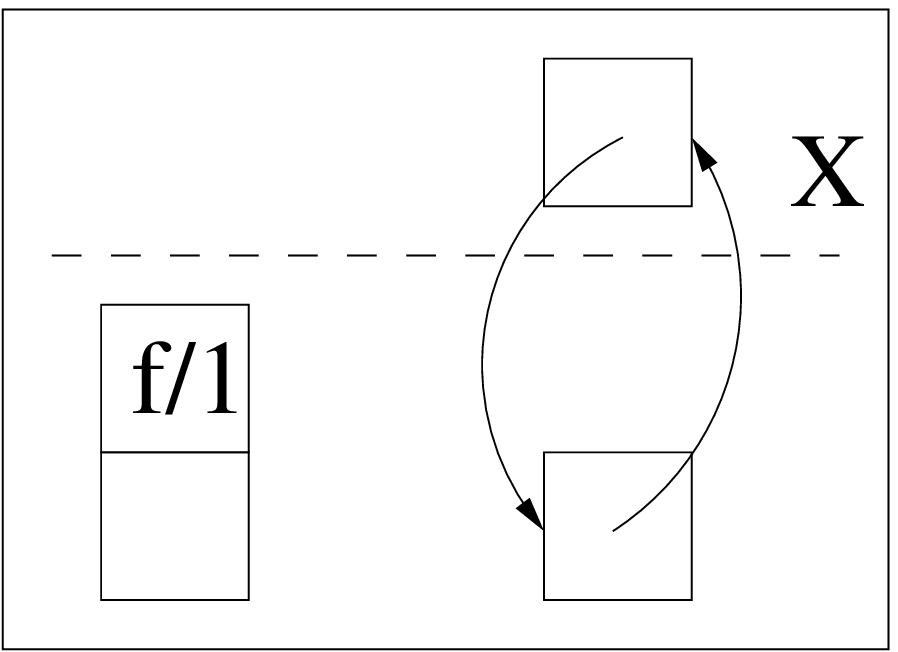,width=.25\textwidth}}}
\subfigure[After.]{{\epsfig{file=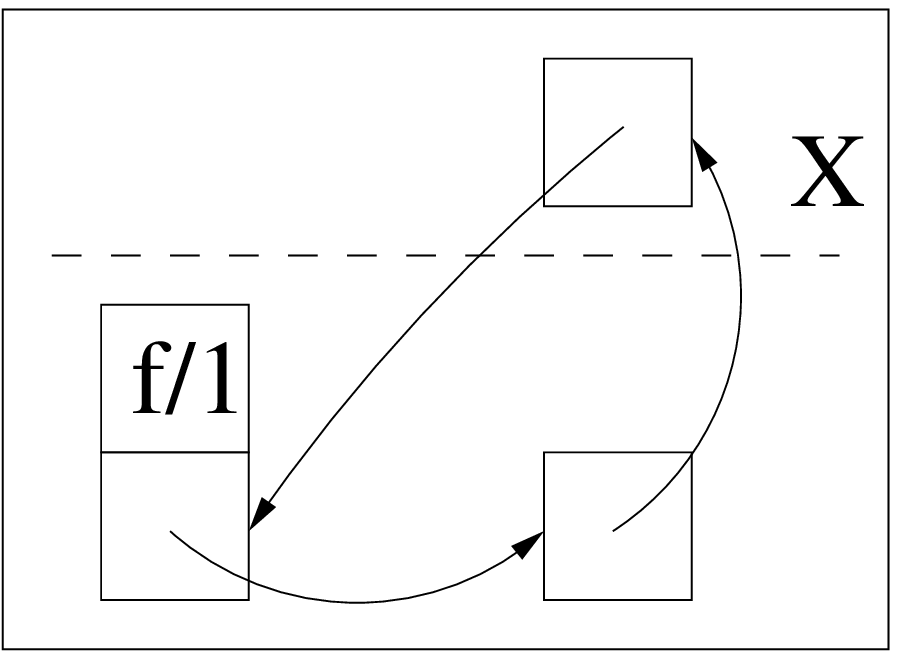,width=.25\textwidth}}}}
\caption{Term construction example: $f(X)$. The dashed line represents a choice-point.} \label{termconstr.fig}
\end{figure}

When a term, e.g. $f(X)$, is constructed with 
$X$ being represented by a PARMA chain,
the argument cell in the structure representation of $f/1$ is inserted in the
chain of $X$ (see \figurename \ref{termconstr.fig}). While $X$
requires shallow trailing, the cell of the term requires no trailing at all
as it is newly created. 

The generalization of this to an $n$-ary variable term 
unification is as follows. 
If all arguments are deep trailed, then $Y$ becomes deep trailed 
and the arguments remain deep trailed.
Otherwise, $Y$ and all its arguments become shallow trailed 
(since each argument is at least shallow trailed by the operation).
Note that if at least one argument was deep trailed, and since each argument shares with
$Y$ after the unification, we must apply $consist$ to maintain the information consistent.

Formally, let $l_1=(s_1,d_1)$ be the pre-description of the unification,
$x$ be the set of variables $\{X_1,\ldots,X_n\}$ and $g_2$ the set of
ground variables after the unification.  Its post-description $l_2$ can be
obtained as:
\[ l_2 = \left\{\begin{array}{l@{~~~}l}
(s_1, d_1 \cup \{Y\}) &  x \subseteq d_1 \\
consist_2(remove\_ground((s_1 \cup x \cup \{Y\}, d_1 \setminus x),g_2)) & \mbox{otherwise}
\end{array}\right.\]

The worst case time complexity is $\mathcal{O}(n^{3})$. This definition
can be combined with the previous one for the overall definition
of variable--term unification. The implementation can be more
efficient, but the complexity will still be $\mathcal{O}(n^{3})$.
        
                \paragraph{Predicate call: $q(X_1 \ldots X_n)$.}

Let $l_1$ be the pre-description of the predicate call and $x$ the set of
variables $\{X_1,\ldots,X_n\}$. The first step
will be to project $l_1$ onto $x$ resulting in description $l_{proj}$.
Note that onto-projection is trivially defined as:
        \[ onto\_proj(l,v) = (s \cap v, d \cap v) \] 
The second step consists in extending $l_{proj}$ onto the set of variables
local to the predicate call. Since these variables are known to be new (and
thus they do not appear in $Old_1$), the extension operation in our domain is
trivially defined as the identity. Thus, from now on we will simply disregard
the extension steps required by HAL's framework.

Let $l_{answer}$ be the answer
description resulting from analyzing the predicate's definition for calling
description $l_{proj}$.  We will assume
that the set $v$ of variables local to $q/n$ has already been
projected out from $l_{answer}$, where out-projection is identical to 
$remove\_ground$, 
which has time complexity $\mathcal{O}(n)$. 

In order to obtain the post-description, we will 
make use of the determinism information. 
Thus, the post-description $l_2$ can be derived by combining the $l_{answer}$
and $l_1$, using the determinism of the predicate call as follows:

\begin{itemize}

\item
If the predicate has determinism \texttt{multi} or \texttt{nondet}
(which can have more than one answer), 
then all variables not in $x$ become not trailed by the (possible)
introduction of a new choice point.
Hence, $l_2$ is equal to $l_{answer}$ except for the fact that we 
have to apply the $consist$
function in order to take into account the changes in sharing involving
variables not in $x$.  

\item
Otherwise, we know the trailing state of variables
in $l_1$ is unchanged except by possibly new introduced sharing.
Thus, $l_2$ is the result of combining $l_{answer}$ and $l_1$ as follows: the
trailing state of variables in $x$ is taken from $l_{answer}$, while that
of other variables is taken from $l_1$. Any deep trailed variables that
share with non-deep trailed variables must, of course, become shallow trailed.

\end{itemize}

Formalized, the combination\footnote{Note that the combination
is not the meet of the two descriptions. It is the 
``specialized combination'' introduced in \cite{diff} which assumes that
$l_{answer}$ contains the most accurate information about the variables in
$x$, the role of the combination being just to propagate this information
to the rest of variables in the clause.} function is defined as:
\[\begin{array}{rcl} l_2 &=& comb(l_1, l_{answer}) \\
                         &=& \left\{\begin{array}{l@{~~~}l}
consist_2(((s_1 \setminus x) \cup s_{answer}, (d_1 \setminus x) \cup
d_{answer})) &  det(q) \leq 1\\
consist_2(l_{answer}) &  \text{otherwise} \\
\end{array}\right.\end{array}\]

Obviously, the complexity is $\mathcal{O}(n^{3})$ because of $consist$. 

\begin{example}
Assume that the call \texttt{q(X)} has pre-description $(\{X,Y\},\emptyset)$
and the predicate \texttt{q/1} has answer description $(\{X\},\emptyset)$. The
post-description of the call depends on the determinism of the predicate.
If the predicate \texttt{q/1} 
has at most one solution, the post-description will
be $((\{X,Y\} \setminus \{X\}) \cup \{X\}, (\emptyset \setminus \{X\}) \cup
\emptyset) = (\{X,Y\},\emptyset)$. Otherwise the post-description will be
equal to the answer description, $(\{X\},\emptyset)$.
\end{example}

                \paragraph{Disjunction: $(G_1 ; G_2; \ldots ; G_n)$.}

Disjunction is the reason why trailing becomes necessary. As mentioned
before, trailing might be needed for all variables which were already
old before the disjunction. Thus, let $l_0$ be the
pre-description of the entire disjunction. Then, $\top$ will be the
pre-description of each $G_i$ except for $G_n$ whose pre-description is 
simply $l_0$ (since the disjunction
implies no backtracking over the last branch).

Let $l_i=(s_i,d_i), 1 \leq i \leq n$ be the
post-description of goal $G_i$. We will again assume
that the set $v_i$ of variables local to each $G_i$ has already been
projected out from $l_i$.  The end result $l_{n+1}$ of
the disjunction is the least upper bound (lub) of all
branches,\footnote{Note that this is not the lub of the \texttt{notrail} domain
alone, but that of the product domain which includes sharing (and
groundness) information.} 
which is defined as:
\[l_1 \sqcup \ldots \sqcup l_n = consist_{n+1}(remove\_ground((s,d),g_{n+1}))\]
where
\[\begin{array}{rclcl}
s &=& (s'_1 \cap \ldots \cap s'_n) \setminus d  \\
d &=& (d'_1 \cap \ldots \cap d'_n)  \\
s_i' &=&s_i \cup d_i'  \\ 
d_i' &=&d_i \cup g_i \\
\end{array}\]

Intuitively, all variables which are deep trailed in all descriptions are
ensured to remain deep trailed; all variables which are 
trailed in all descriptions but have not always been deep trailed (i.e.,
are not in $d$) are ensured to have already been (at least) shallow
trailed. Note that variables which are known to be ground in all descriptions
(those in $g_{n+1}$) are eliminated. This is consistent
with the view that only old variables are represented by the
descriptions and avoids adding overhead to the abstract operations.

HAL also includes switches, which are disjunctions where the
compiler has detected that only one branch needs to be executed.
Switches are treated identically to disjunctions except for the fact that
the pre-description for each $G_i$ is $l_0$ rather than $\top$.

\begin{example}\label{ex:disj}
Let $l_0=(\emptyset, \{X,Y,Z\})$ be the pre-description  of 
the code fragment:
\begin{Verbatim}[xleftmargin=30mm]
( A = a, X = Y ; A = b, X = f(Y, Z) )               
\end{Verbatim}
Let us assume there is no sharing at that program point. 
Assuming that $A$ is old, then this is simply a disjunction.
Then, the
pre-descriptions of the first branch is $(\emptyset, \emptyset)$, the
$\top$ element of our domain. The pre-description of the second branch
is $(\emptyset,\{X,Y,Z\})$, i.e., since this is the last branch in the
disjunction, its pre-description is identical to the pre-description of the
entire disjunction. Their post-descriptions are $(\{X,Y\},\emptyset)$ and
$(\emptyset,\{X,Y,Z\})$, respectively. Finally, the lub of the two
post-descriptions results in $(\{X,Y\},\emptyset)$.

Now assume $A$ is ground. Then this code fragment is a switch
on $A$. The pre-description for the first branch becomes
$(\emptyset,\{X,Y,Z\})$ and the post description is the same. Finally the
lub of the two post-descriptions for the two branches is
$(\emptyset,\{X,Y,Z\})$.
\end{example}

The time complexity of the joining of the branches is simply that of the
lub operator ($\mathcal{O}(n^3)$) for a fixed maximum number of branches,
and it is completely dominated by the $consist_{n+1}$ function.

\paragraph{If-then-else: $I\ \rightarrow \ T\ ; E\ $.}

Although the if-then-else could be treated as $(I,T;E)$, this is rather
inaccurate since (as in the case of switches) only one branch will ever be
executed and, thus, there is no backtracking between the two branches.

Hence, we can do better if no old variable that exists before the if-then-else
is bound or aliased, i.e. possibly requiring trailing and backtracking if
the condition fails.  This is not a harsh restriction, since it is ensured
whenever the if-condition is used in a logical way, i.e., it simply inspects
existing variables and does not change any non-local variable. However,
in general it is not possible to statically determine this property. Instead
a safe approximation is used: the if-then-else is treated as $(I,T;E)$ if
the condition contains any pre-existing old variables, otherwise the following
stronger treatment is used.

Let $l_1$ be the pre-description to the if-then-else. Then $l_1$ will also be
the pre-description to both $I$ and $E$. Let $l_I$ be the
post-description obtained for $I$. Then $l_I$ will also be the pre-description
of $T$. Finally, let $l_T$ and $l_E$ be the post-descriptions obtained for
$T$ and $E$, respectively. Then, the post-description for the if-then-else
can be obtained as the lub $l_T \sqcup l_E$.

The time complexity of the joining of the 
branches is again $\mathcal{O}(n^{3})$, just like
the operation over the disjunction.

\begin{example}
Let $l_0=(\emptyset, \emptyset)$ be the pre-description  of 
the following if-then-else where $N$ is known to be ground:
\begin{Verbatim}[xleftmargin=30mm]
( N = 1 -> X = Y  ; X = f(Y, Z) )               
\end{Verbatim}
Assume no variables share before the if-then-else. 
Then, $l_0$ is equal to the pre-description 
of both the then- and else-branch. The post-de-\linebreak scription
of the then-branch is $(\{X,Y\},\emptyset)$ and that of the else-branch
is \linebreak $(\{X,Y,Z\},\emptyset)$. The post-description finally is obtained
as their lub: $(\{X,Y\},\emptyset)$.

If the pre-description was $l_0=(\emptyset, \{X,Y,Z\})$
as in Example~\ref{ex:disj}, then the post-description would be 
$(\emptyset, \{X,Y,Z\})$, since no additional trailing will be required.
\end{example}

         \paragraph{Higher-order term construction: $Y = p(X_1,\ldots,X_n)$.}

This involves the creation of a partially evaluated predicate, i.e., we are
assuming there is a predicate with name $p$ and arity equal or higher than
$n$ for which the higher-order construct $Y$ is being created. In HAL, $Y$
is required to be new. Also, it is often too difficult or even
impossible to know whether $Y$ will be actually called or not and, if so,
where. Thus, HAL follows a conservative approach 
and requires that the instantiation of the ``captured'' arguments
(i.e., $X_1, \ldots, X_n$) remain unchanged after calling $Y$.
It also guarantees (through type and mode checking) that no higher-order terms are ever unified.

The above requirements allow us to follow a simple (although conservative)
approach: Only after a call to $Y$ will the trailing of the captured variables
be affected. If the call to $Y$ might have more than one solution and thus may
involve backtracking, then the involved variables will be treated safely in
the analysis at the call location if they are still statically live there.

If the call to $Y$ does not involve backtracking but does involve unifications,
then trailing information might not be inferred correctly at the call
location. This is because the captured variables are generally not known
at the call location. To keep the trailing information safe, any potential
unifications have to be accounted for in the higher-order unification. Since
the construction of the higher-order term involves no backtracking and all
unifications leave the variables they involve at least shallow trailed,
it is sufficient to demote all captured deep trailed variables to shallow
trailed status, together with all sharing deep trailed variables.

Formally, let $l_1=(s_1,d_1)$ be the pre-description of the higher-order
term construction and $x$ be the set of variables $\{X_1,\ldots,X_n\}$. 
Then its post-description $l_2$ can be obtained with a time complexity of $\mathcal{O}(n^{3})$ as:
\[ l_2 = \left\{\begin{array}{l@{~~~}l}
consist_2((s_1 \cup (x \cap d_1),d_1 \setminus x)) &  x \cap d_1 \not = \emptyset \\
l_1 & \mbox{otherwise} \\
\end{array}\right. \]
                
                \paragraph{Higher-order call: $call(P,X_1,\ldots,X_n)$.}

The exact impact of a higher-order call is difficult to determine in
general.  Fortunately, even if the exact predicate associated to variable
$P$ is unknown, the HAL compiler still knows its determinism. This can help
us improve accuracy. If the predicate might have more than one solution,
all variables must become not trailed. Since the
called predicate is typically unknown, no answer description is available
to improve accuracy.

Otherwise, the worst that can happen is that the deep trailed 
arguments of the call become shallow trailed. So in the post-description
we move all deep trailed arguments to the set of shallow trailed variables,
together with all variables they share with. Recall that for this case
the captured variables have already been taken care of 
when constructing the higher-order term.

The sequence of steps is much the same as that for the predicate call.
First, we project the pre-description $l_1$ onto the set $x$ of variables
$\{X_1,\ldots,X_n\}$, resulting in $l_{proj}$. Next, the answer description
$l_{answer}$ of the higher-order call is computed as indicated above:
\[l_{answer} = \left\{\begin{array}{l@{~~~}l}
(s \cup d, \emptyset)& det(P)  \leq 1 \\
(\emptyset, \emptyset)&  \text{otherwise}\\
\end{array}\right.\]
The combination of $l_{answer}$ and $l_1$ is computed to obtain the 
post-description $l_2$.

\section{Trailing Optimization}\label{sec:ntdom:opt}

The optimization phase consists of 
deciding for each unification in the body
of a clause which variables need to be trailed. This decision is based on
the pre-description of the unification, inferred by the trailing analysis.
If some variables do not need to be trailed, the general unification
predicate is replaced with a variant that does not trail those
particular variables.  Thus, we will need a different variant for each
possible combination of variables that do and do not need to be trailed.
\begin{itemize}
\item
For the unification of two unbound variables, trailing is omitted for either
variable if it is shallow trailed or deep trailed in the pre-description.
\item
For the binding of an unbound variable $X$, 
trailing of $X$ is omitted if it
is deep trailed in the pre-description. 
\item In the construction of a term containing an
old unbound variable $X$, trailing of $X$ is omitted if 
$X$ is either shallow or deep trailed in the pre-description.
\item
For the unification of two bound variables, the trailing for chains in the
structure of either is omitted if it is deep trailed in the
pre-description.
\end{itemize}
Often it is not known at compile time whether a variable is bound or not, 
so a general variable-variable unification predicate is required that performs
run-time boundness tests before selecting the appropriate kind of unification.
Various optimized variants of this general predicate are needed as well.

Experimental results for the analysis are presented in Section~\ref{results}.

\section{The improved trailing scheme}\label{improved_trailing}

Let us now present a trailing scheme which is more sophisticated than the
classic PARMA value trailing discussed in Section~\ref{background}. We will
start by considering the improvements that apply to each kind of
unification (variable--variable and variable--nonvariable) and finish by
showing how to combine them.

Our modified scheme
must be able to apply different untrail operations depending on the 
kinds of trailing that was performed. A simple tagging scheme (explained in detail in Section~\ref{combiningtrailings}) is used to
indicate the kind of untrailing required in each case. 

\subsection{Variable--variable unification: swap trailing}

In the classic scheme the value trailing of both cells takes up four trail
stack slots (two for the addresses of each variable plus another two for
their contents) when trailing is unconditional. Undoing such
variable--variable unification consists of simply restoring the old values
of the cells separately. However, there is a more economic inverse
operation that undoes the swapping that happened during unification: simply
swapping back. This swapping only requires the addresses of the involved
cells and not their respective old contents. We introduce a new kind of
trailing named {\em swap trailing} which exploits this and also the
corresponding untrailing operation. Swap trailing is defined by the
following code:
\begin{Verbatim}
         swaptrail(p, q) {
               *(tr++) = p;                      
               *(tr++) = set_tag(q,SWAP_TRAIL);
         }          
\end{Verbatim}
where \texttt{p} and \texttt{q} are the addresses of the two cells, 
\texttt{tr} is a pointer to the top of the trailing stack, \texttt{SWAP\_TRAIL}
is a tag, and the function 
\texttt{set\_tag(c,t)} tags cell \texttt{c} with tag \texttt{t}. Note that swap
trailing only consumes two slots in the trail stack, as opposed to the four 
used by (unconditional) value trailing in the classical scheme. The untrail operation
for swap trailing is:
\begin{Verbatim}
         untrail_swaptrail() {
               q = untag(*(--tr)); /* recover address q */
               p = *(--tr);        /* recover address p */
               tmp = *q;        
               *q = *p;            /* swap contents of p with q */
               *p = tmp;
         }
\end{Verbatim}
The above improvement assumes that both cells are unconditionally
trailed. If conditional value trailing is available, the classic scheme
would either consume zero, two or four slots if respectively none, only one
or both variables are older than the most recent choice point. Swap
trailing can only be used in conjunction with conditional trailing to
replace the four slot case, with value trailing still needed for the two
slot case. As a result the code for conditional variable--variable trailing
looks like:
\begin{Verbatim}
         cond_varvartrail(p, q, bh) {
               if (is_older(p,bh)) {               
                 if (is_older(q,bh)) {             
                    swaptrail(p,q);  /* trail both using swaptrail */     
                 } else {                  
                    valuetrail(p);  /* only trail p */   
                 }                         
               } else if (is_older(q,bh)) {        
                  valuetrail(q);    /* only trail q */   
               }
         } 
\end{Verbatim}
It is important to note that the potential gain in space on the trail obtained
by the above operations comes at a cost
in execution time (more run-time operations are needed) and that the gain in space is not guaranteed.

\subsection{Variable--nonvariable unification: chain trailing}

As seen before, variable-nonvariable unification pulls the entire chain of
the variable apart by setting every cell in the chain to the
nonvariable. In the case of classic value trailing, every address of a cell
is stored twice: once as the address of a cell and once as the contents of
the predecessor cell. This means that there is quite some redundancy. The
obvious improvement is to store each address only once. We name this {\em
chain trailing}. Because the length of the chain is not known, a marker is
needed to indicate, for the untrailing operation, where chain trailing
ends. The last entry of the chain encountered during untrailing, is the
first one actually trailed.  We use the \texttt{CHAIN\_END} tag to mark
this entry.

The last address put on the trail is tagged with \texttt{CHAIN\_BEGIN} to
indicate the kind of trailing. For chains of length one, the last and first
cell coincide. The \texttt{CHAIN\_END} tag is used to mark this single address.

Chain trailing is defined by the code:
\begin{Verbatim} 
    chaintrail(p) {          
       start = p;                      
       *(tr++) = set_tag(p,CHAIN_END); 
       p = *p;
       only_one = TRUE;                         
       while (p != start) {       /*trail each cell address*/
          only_one = FALSE;                
          *(tr++) = p;                 
          p = *p;
       }
       if (!only_one) {           /* if more than one cell */
          last = tr - 1;          /* tag last one as CHAIN_BEGIN*/
          *last = set_tag(*last,CHAIN_BEGIN);
       }
    }
\end{Verbatim}

The untrail operation for reconstructing the chain is straightforward: it
dispatches to the appropriate untrailing action depending on the tag
of the first cell encountered during untrailing. If this is
\texttt{CHAIN\_BEGIN}, meaning $n \ge 1$, the corresponding code is:
\begin{Verbatim}
         untrail_chaintrail() {
            head = untag(*(--tr));
            previous = head;
            current = *(--tr);
            while (get_tag(current) != CHAIN_END) {
              *current = previous;
              previous = current;
              current = *(--tr);
            }
            current = untag(current);
            *current = previous;
            *head = current;
         }
\end{Verbatim}
If the first tag is \texttt{CHAIN\_END}, then $n = 1$ and
the code for untrailing is:
\begin{Verbatim}
         untrail_shortchain() {
            cell = untag(*(--tr));
            *cell = cell;
         }
\end{Verbatim}

\begin{example}
Consider the trailing that occurs using
the improved scheme for the goal
\texttt{X = Y, Z = W, X = Z, X = a} from Example~\ref{ex:basictrail}.
The first unification is a swaptrail, trailing \texttt{X} and \texttt{Y}, similarly
the second unification swaptrails \texttt{Z} and \texttt{W} and the third unification
swap trails \texttt{X} and \texttt{Z}. 
Finally the last unification chain trails \texttt{X}.
The resulting trail looks like:
\begin{center}
\fbox{\texttt{X}} \fbox{\texttt{Y}}$^{\texttt{sw}}$ 
\fbox{\texttt{Z}} \fbox{\texttt{W}}$^{\texttt{sw}}$ 
\fbox{\texttt{X}} \fbox{\texttt{Z}}$^{\texttt{sw}}$ 
\fbox{\texttt{X}}$^{\texttt{ce}}$ \fbox{\texttt{W}} \fbox{\texttt{Z}} 
\fbox{\texttt{Y}}$^{\texttt{cb}}$ 
\end{center}
where we use superscripts
\texttt{sw}, \texttt{cb} and \texttt{ce} to 
represent the \texttt{SWAP\_TRAIL}, \texttt{CHAIN\_BEGIN} and
\texttt{CHAIN\_END} tags respectively.
This uses 10 trail entries compared to the 24 entries in
Examples~\ref{ex:basictrail} and~\ref{ex:more} 
\end{example}

The above improvement assumes that all cells are unconditionally trailed.
Let us assume that the chain consists of $n$ cells, $k$ of which are older
than the most recent choice point.  If conditional trailing is available
and $2*k < n$, our unconditional chain trailing will consume more space
than the classic conditional value trailing. Fortunately, a conditional
variant of chain trailing is also possible:
\begin{Verbatim} 
    cond_chaintrail(p, bh) {              
       start = p;                               
       first = TRUE;
       only_one = TRUE;                            
       do {                                     
          if (is_older(p,bh))    /* trail each older cell in chain*/
             if (first) {                       
                *(tr++) = set_tag(p,CHAIN_END); /*tag if first*/
                first = FALSE;                  
             } else {
                only_one = FALSE;                           
                *(tr++) = p;                    
             }                                  
          p = *p;                               
       } while (p != start);
       if (!only_one) {         /* if more than one older cell */
          last = tr - 1;        /* tag last one as CHAIN_BEGIN*/
          *last = set_tag(*last,CHAIN_BEGIN);
       }
    } 
\end{Verbatim}
This conditional variant uses only $k$ slots of the stack trail, so it is
clearly an improvement over conditional value trailing whenever $k > 0$. 

Note that the untrail operation used is the same as for the unconditional
chain trailing. This might look wrong at first since the {\tt
cond\_chaintrail} might not trail all cells in the chain. However, this is
simply exploiting the fact that the objective of trailing is to be able to
reconstruct the bindings that existed at the creation time of a choice
point. Thus, the final state of younger cells and the state of any cell
during the intermediate steps of untrailing are irrelevant. In fact, the
more general -- and better with respect to stack trail consumption --
principle behind this is that only the old cells (older than the most
recent choice point) in the chain pointing to other old cells have to be
trailed (an old cell must have been made to point to a new cell after the last choice-point). 
The kind of trailing suitable for this
insight is a special kind of value trailing, where the successive equal
slots on the trail stack are overlapped. The above {\tt cond\_chaintrail}
operation only approximates this, since an implementation would incur an
undue time overhead because of the extra run-time tests needed to test the
age of the successors. Thus, we store the addresses of old cells even if
they neither point to nor are pointed to by old cells.

\begin{figure}[!ht]
\begin{centering}
\subfigure[Initially]{{\epsfig{file=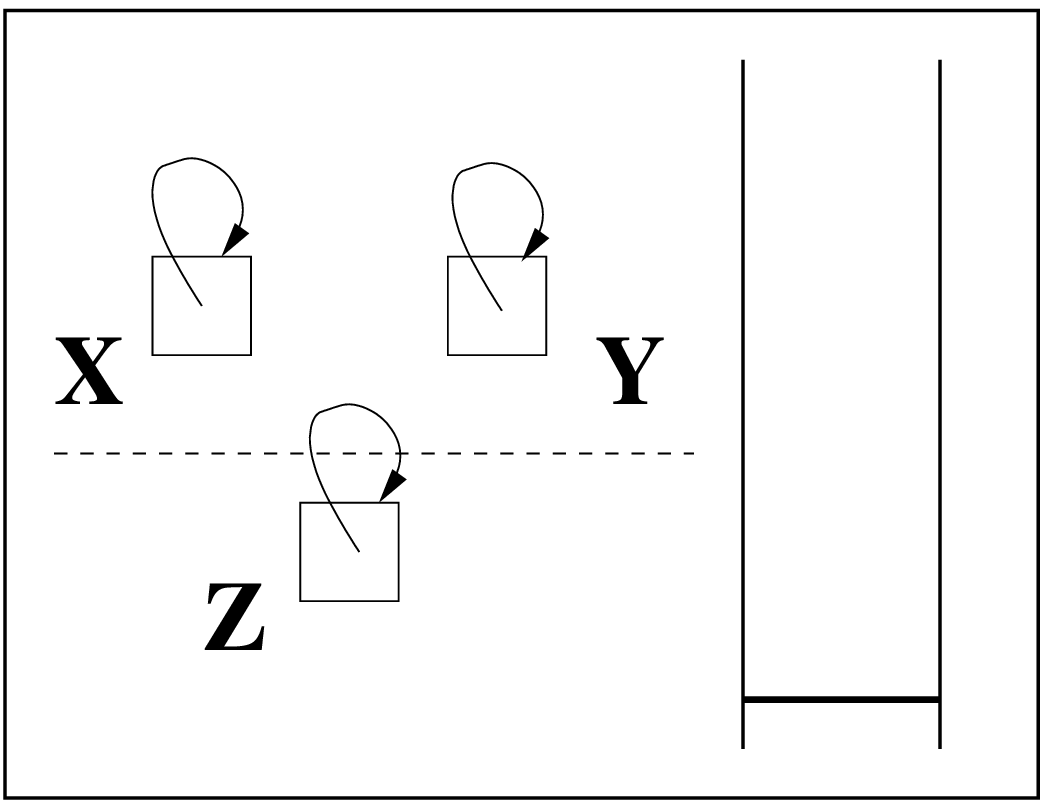,width=.20\textwidth}} \label{initial}}
\subfigure[\texttt{X = Z}]{{\epsfig{file=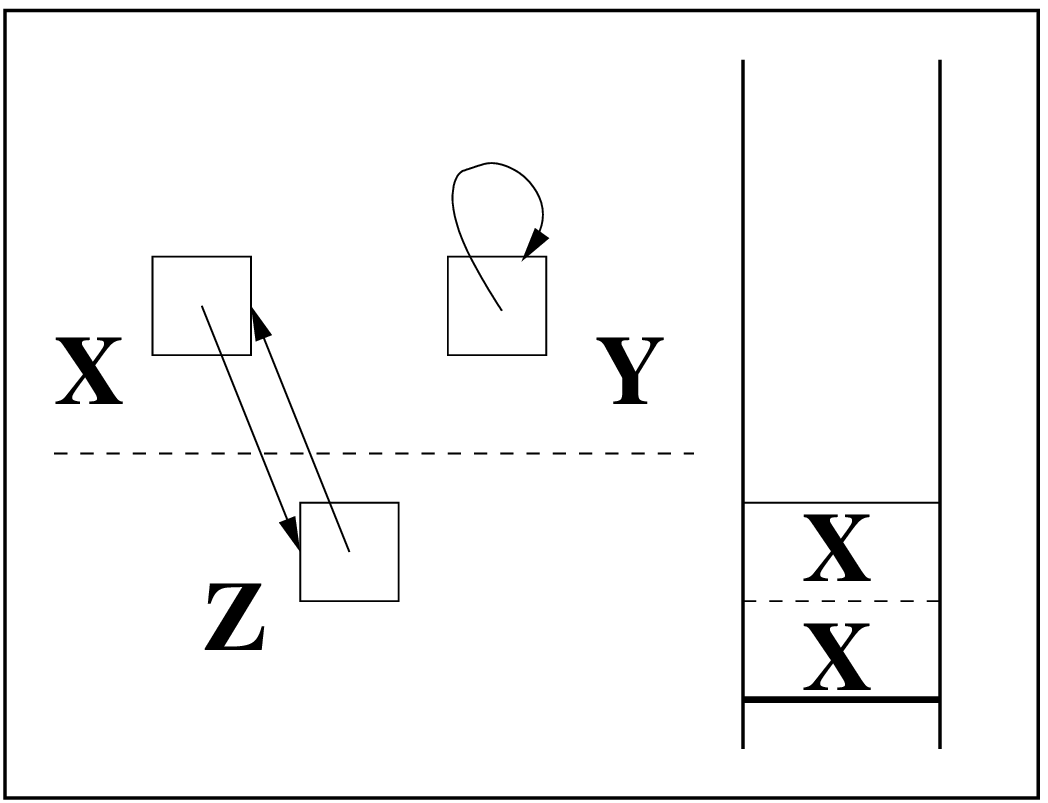,width=.20\textwidth}} \label{step1}}
\subfigure[\texttt{Z = Y}]{{\epsfig{file=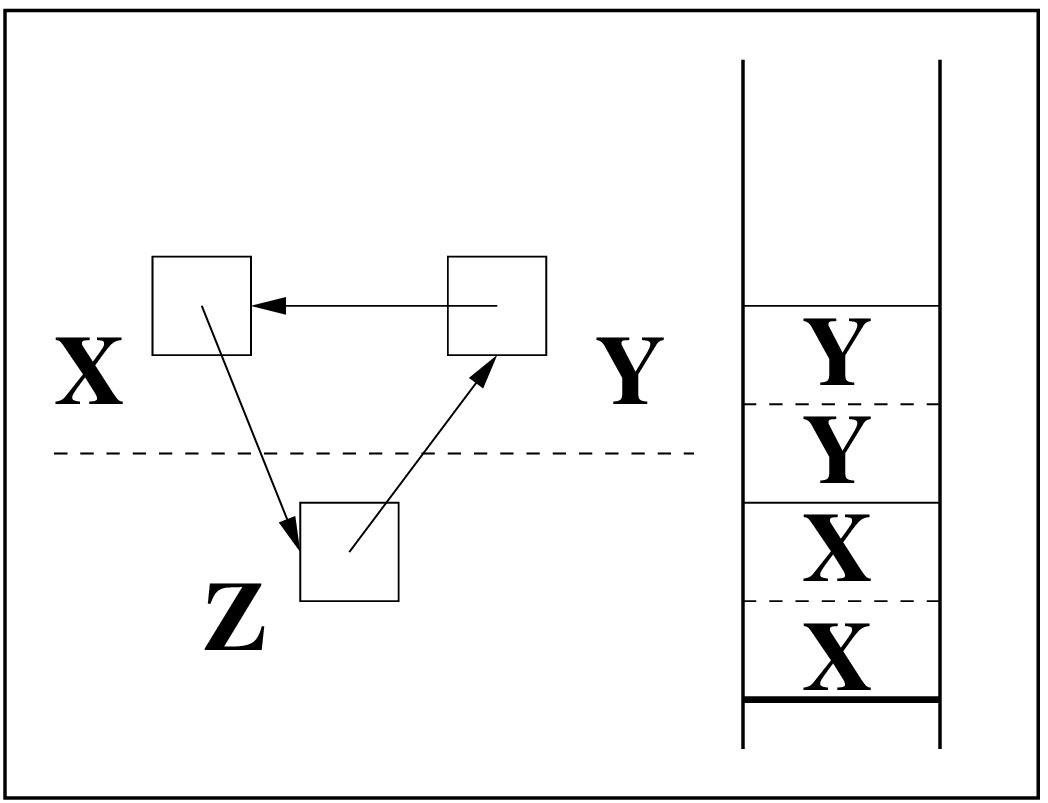,width=.20\textwidth}} \label{step2}}
\subfigure[\texttt{X = a}]{{\epsfig{file=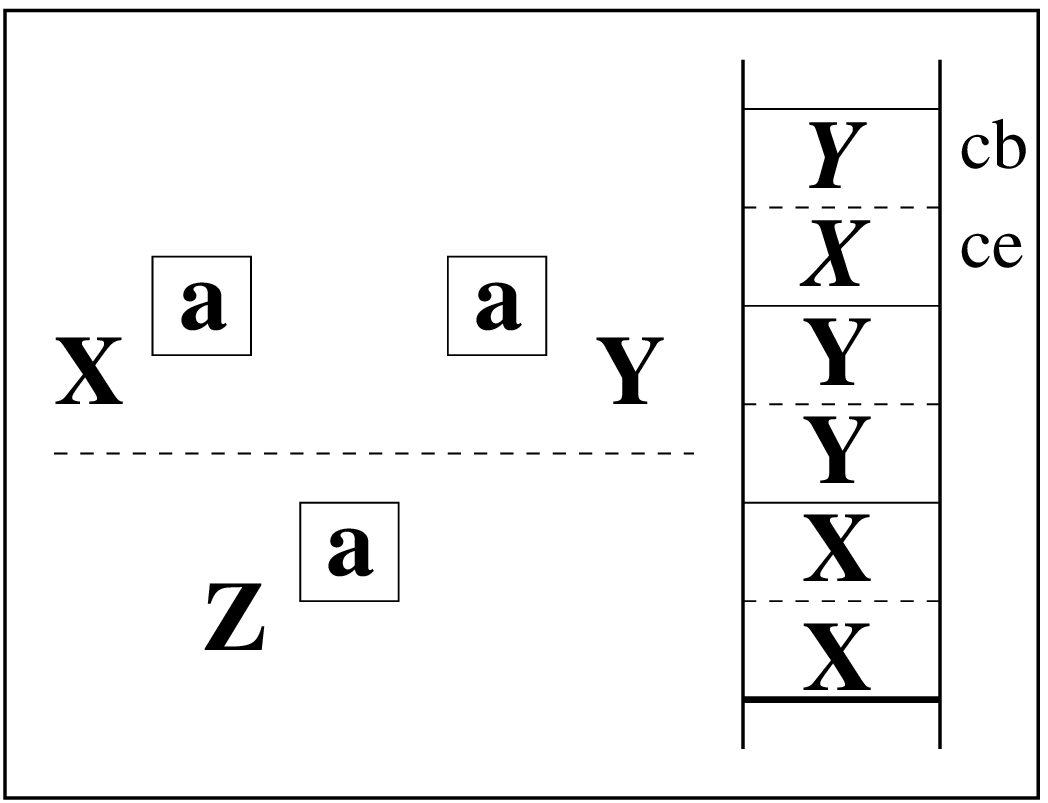,width=.20\textwidth}} \label{step3}}
\subfigure[Untrail \texttt{X = a}]{{\epsfig{file=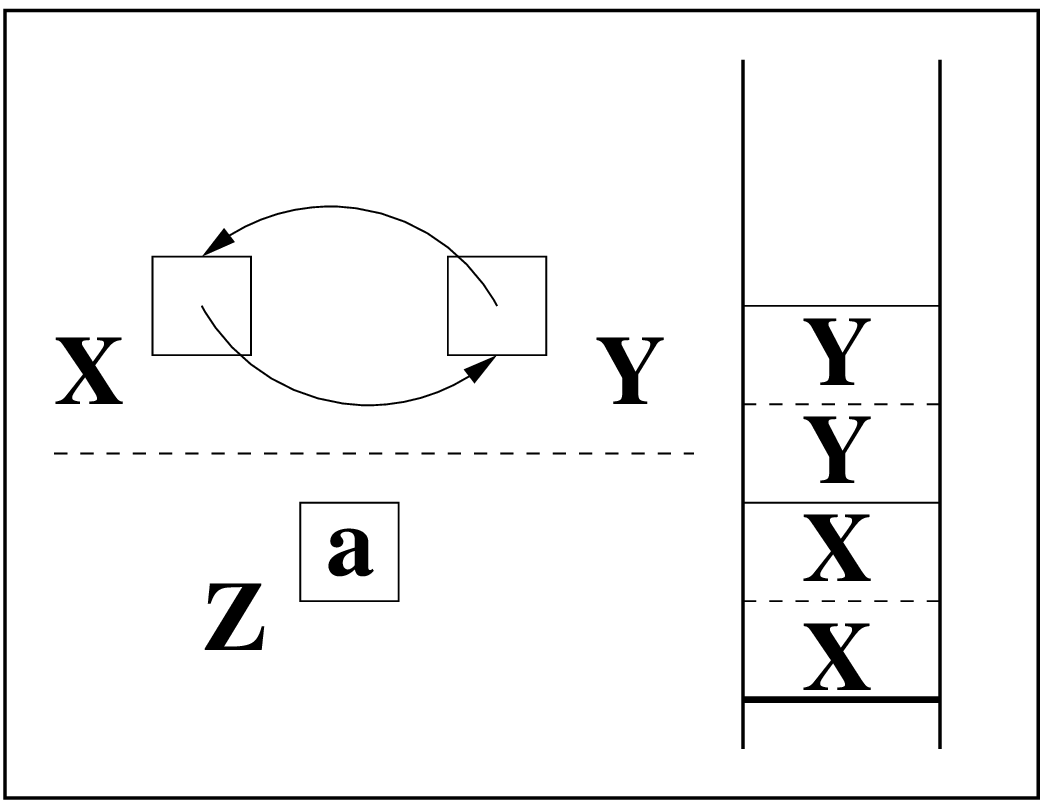,width=.26\textwidth}} \label{step3u}}
\subfigure[Untrail \texttt{Z = Y}]{{\epsfig{file=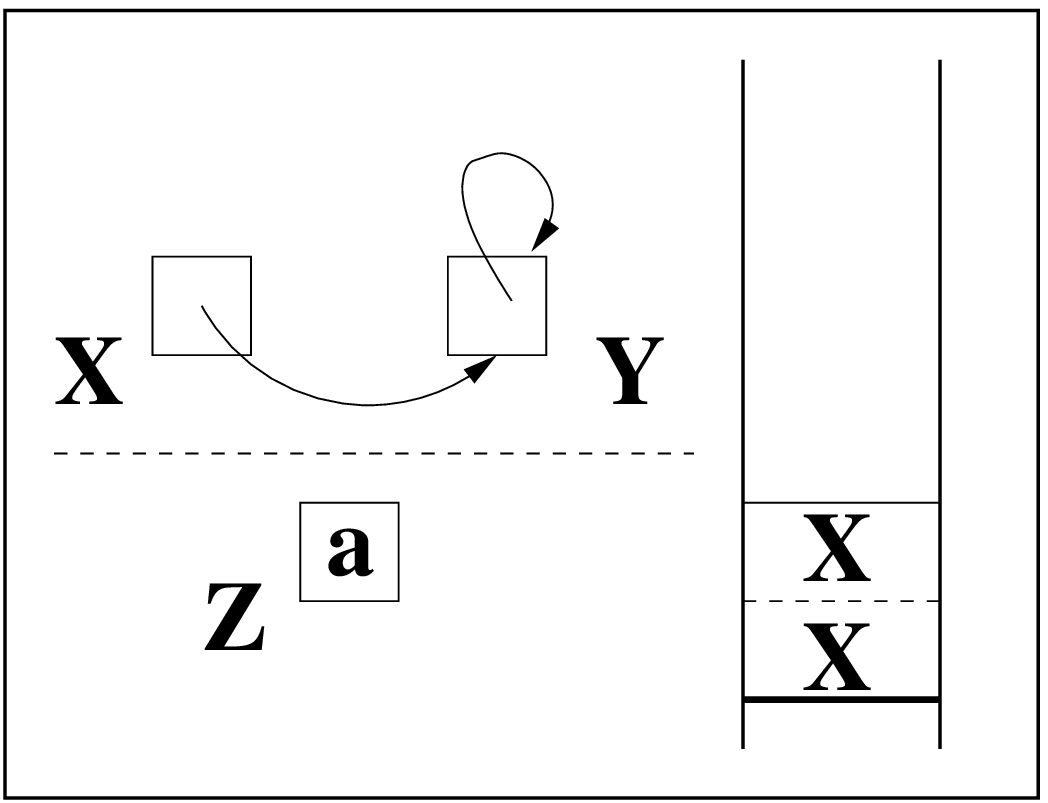,width=.26\textwidth}} \label{step2u}}
\subfigure[Untrail \texttt{X = Z}]{{\epsfig{file=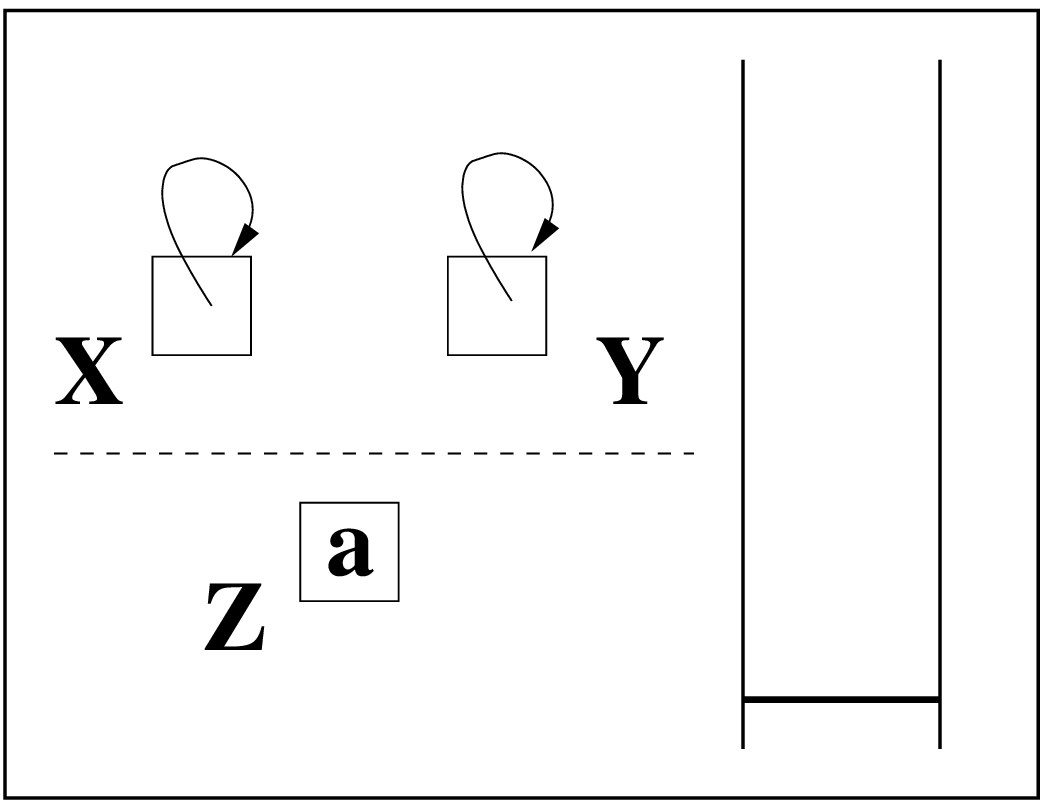,width=.26\textwidth}} \label{step1u}}
\caption{Conditional chain trailing example.} \label{chaintrailfig}
\end{centering}
\end{figure}

\begin{example}\label{ex:cond-safely}
Figure~\ref{chaintrailfig} illustrates with a small example how the above
specified conditional chain trailing, together with previous trailings, safely
restores the state of all variables older than the most recent choice point. 
Consider the following goal \texttt{X = Z, Z = Y, X = a, fail}
and let us assume that both 
\texttt{X} and \texttt{Y} are older than the most recent
choice point, \texttt{Z} is newer, and all three are chains of length 1 as
depicted in Figure~\ref{initial}.  The successive forward steps are shown
in the Figures \ref{step1}, \ref{step2} and \ref{step3}. \texttt{X} is value
trailed during \texttt{X = Z}, as is \texttt{Y} during \texttt{Z = Y}. The
addresses of \texttt{X} and \texttt{Y} are stored on the trail stack with
conditional chain trailing during \texttt{X = a}\footnote{This could be
avoided if \texttt{X} is known to have been trailed already.}.
The \texttt{cb} and \texttt{ce} to the side of the stack trail
entries represent the \texttt{CHAIN\_BEGIN} and
\texttt{CHAIN\_END} tags respectively.

The execution fails immediately after \texttt{X = a}, and backtracks to the
initial state in three steps. First (Figure~\ref{step3u}), the
conditional chain trailing is untrailed, creating a chain of \texttt{X} and
\texttt{Y}. Next (Figure~\ref{step2u}), the value trailing of \texttt{Y}
is undone and finally (Figure~\ref{step1u}), the value trailing of
\texttt{X} is reversed too. The final state corresponds to the initial
state, except for \texttt{Z}, which is still bound to a. However, as
\texttt{Z} did not exist before the most recent choice point, its content
is irrelevant at that point because it is inaccessible and will be
reclaimed from the heap anyway when forward execution resumes.
Note the \emph{illegal} intermediate state illustrated in  Figure~\ref{step2u}
is not important since it only occurs in the middle of untrailing, and
never during execution.
\end{example}

\subsection{Combining the improvements} \label{combiningtrailings}

Let us first consider the combination in the context of the modified unconditional 
trailing scheme of the Mercury back-end of HAL. 
In this context, in addition to
swap and unconditional chain trailing, function trailing is used to
allow custom trailings for constraint solvers. Function trailing stores a
pointer to an untrailing function and to untrailing data. 
Thus, we need four different 
tags to distinguish the different trailing information that can appear on the 
trail. 
Fortunately, there are two tag bits available (because of the aligned
addressing for 32 bit machines).
There is
one constraint on the allocation of the four different tags to the kinds of
trailing: the \texttt{CHAIN\_END} tag should not look the same as the tag
of the intermediate addresses in a chain trail.

The general untrail operation then simply looks like:

\begin{Verbatim}
        untrail(tr_cp) {
           while (tr > tr_cp) {
              switch (get_tag(*tr)) {
                 case FUNCTION_TRAIL:
                    untrail_functiontrail();
                    break;
                 case SWAP_TRAIL:
                    untrail_swaptrail();
                    break;
                 case CHAIN_BEGIN:
                    untrail_chaintrail();
                    break;
                 case CHAIN_END:
                    untrail_shortchain();
              }
           }
        }
\end{Verbatim}
Note that, since we are assuming we are in a modified unconditional
trailing scheme, value trailing is never used. This is because value
trailing is only needed in the modified scheme whenever only one of the two
variables involved in a variable-variable unification is newer than the
most recent choice point, and thus only that one was trailed. Otherwise
swap trailing will be used. Since no conditional trailing is allowed, swap
trailing is always used for variable-variable unifications.

Let us now consider the combination in the context of the modified conditional 
trailing scheme of dProlog. In this context only value, swap
and conditional chain trailing are used. The remarks on the application and
allocation of tags is the same as for the unconditional case and the general
conditional untrail operation looks identical except for the fact that the 
\texttt{FUNCTION\_TRAIL} case is substituted by a \texttt{VALUE\_TRAIL} case,
and the call to \texttt{untrail\_functiontrail()} is substituted by a 
call to \texttt{untrail\_valuetrail()}.

When looking at the value trailings of chains of length one in the
example in the previous section (see Figure~\ref{chaintrailfig}),
there is an obvious trailing alternative in the conditional system
that stores no redundant information: chain trailing. Indeed, if such
a variable would be chain trailed instead of value trailed, only one
instead of two slots would be used on the stack. However, this would
require more run-time tests and we have not implemented this.
\begin{Verbatim}
         value_trail(p) {
	    if (*p == p) /* self pointer */
	       *(tr++) = set_tag(p,CHAIN_END);
	    else
	       *(tr++) = *p;
	       *(tr++) = p;
	 }
\end{Verbatim}

Experimental results for both the conditional and unconditional trailing
scheme are presented in Section~\ref{results}.

\section{Analysis for the improved trailing scheme}\label{analysis-impr}

Trailing analyses heavily depend on the details of the trailing scheme. The
analysis presented in Section \ref{analysis} was defined for the classic
PARMA trailing scheme. In this section we present the modifications needed
by that analysis in order to be applied to our improved trailing scheme. As
we will see, the improved scheme gives rise to fewer opportunities for
trail savings.

\subsection{Unnecessary trailing in the improved trailing scheme}

The main difference between the two schemes in terms of unnecessary
trailing appears when considering cells that have been trailed since the
most recent choice-point. In the case of value- and chain-trailing, these
cells do not need to be trailed again since the information stored the
first time allows us to reconstruct the state right before the
choice-point.\footnote{This is assuming that the semantics of function
trailing is such that it does not rely on the intermediate state of any
Herbrand variable during untrailing.} As we will see 
later in the experimental evaluation, this allows our previous analysis
to detect many spurious trailings.

In the case of swap trailing, however, cells need to be trailed even if
they have already been trailed since the most recent choice-point. This is
because swap trailing is an incremental kind of trailing (the
content of the cells is not stored during the trailing, but only the
incremental change) and thus relies on
future trailings for proper untrailing of cells. As a result, during the untrailing process in our
improved scheme, all later chain and swap trailings have to be undone
before the swap trailing can be untrailed correctly. Thus, there is no
opportunity here to avoid future trailings  between
two choice points, after the first trailing has been performed. Let us illustrate this
with a counterexample.

\bigskip
\begin{cxmpl}
Let us not trail variables a second time
between two choice points. Consider then the following code:
\begin{Verbatim}
         X = Y, Z = W, X = Z, fail
\end{Verbatim}
where all variables are older than the most recent choice point and, 
initially, they  are represented as chains of length one, as depicted in 
Figure~\ref{ex2initial}. In the first two steps the
four variables are aliased and swap trailed pairwise, creating two chains
of length two (see Figure~\ref{ex2step1}). The \texttt{s}'s represent
\texttt{SWAP\_TRAIL} tags.

\begin{figure}[h!]
\begin{centering}
\subfigure[Initially]{{\epsfig{file=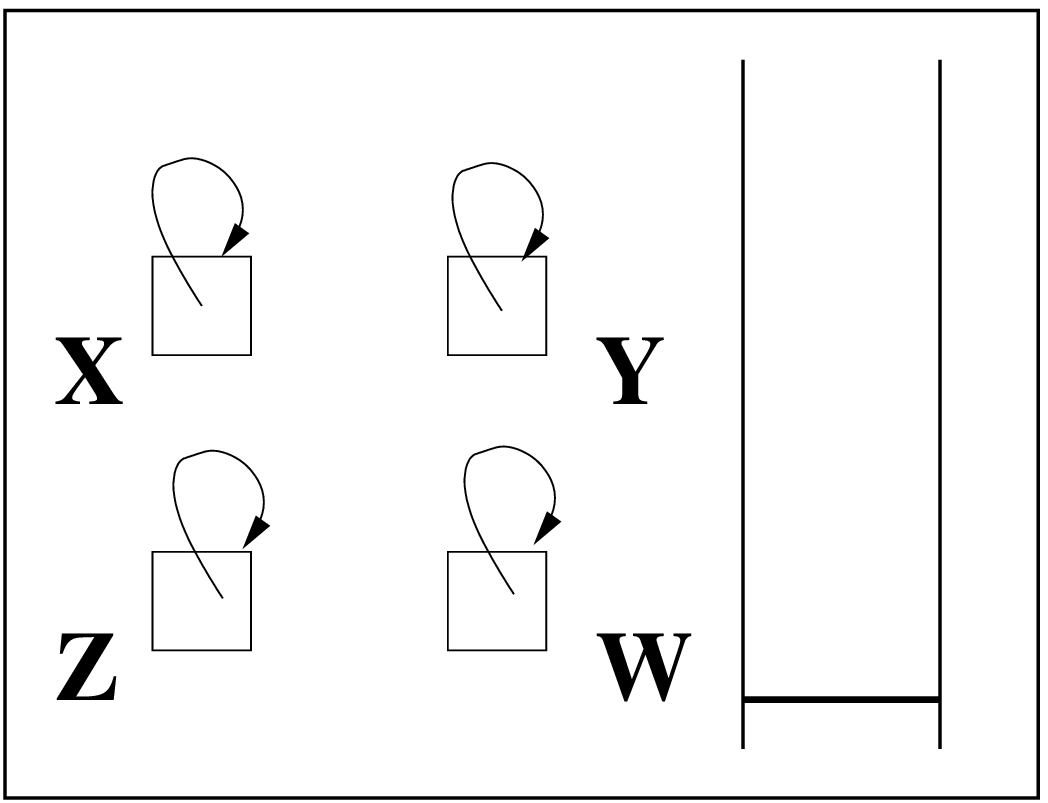,width=.20\textwidth}} \label{ex2initial}}
\subfigure[\texttt{\mbox{X = Y,} \mbox{Z = W}}]{{\epsfig{file=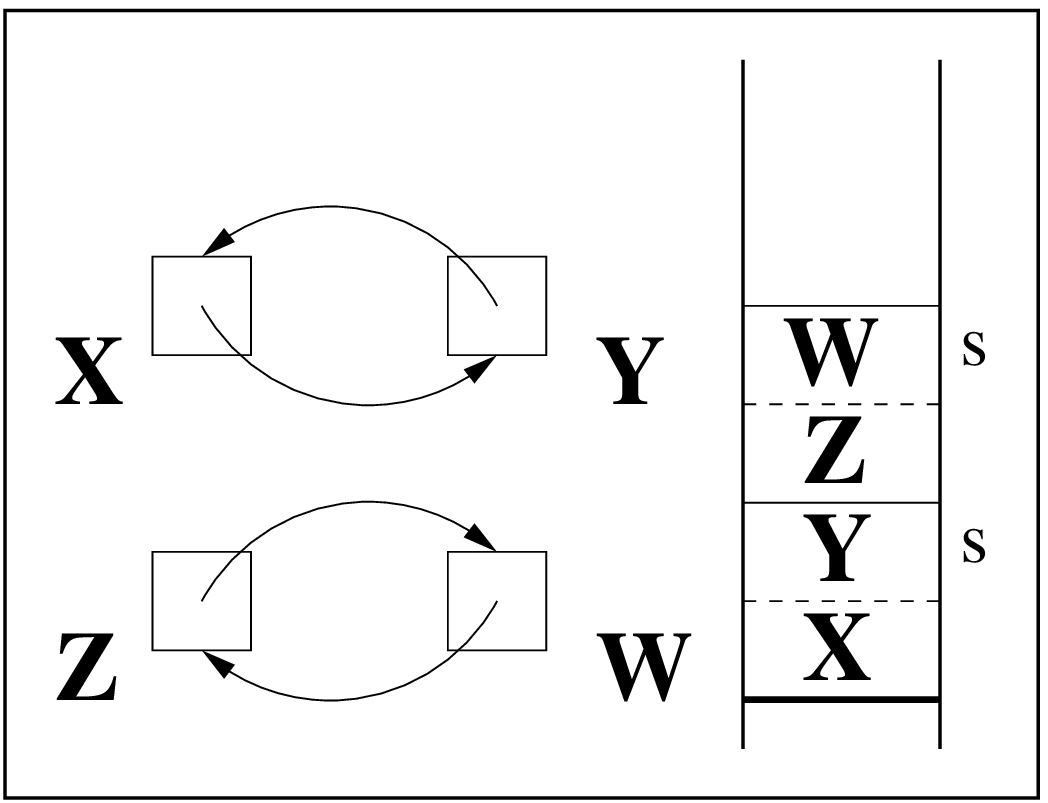,width=.20\textwidth}} \label{ex2step1}}
\subfigure[\texttt{X = Z}]{{\epsfig{file=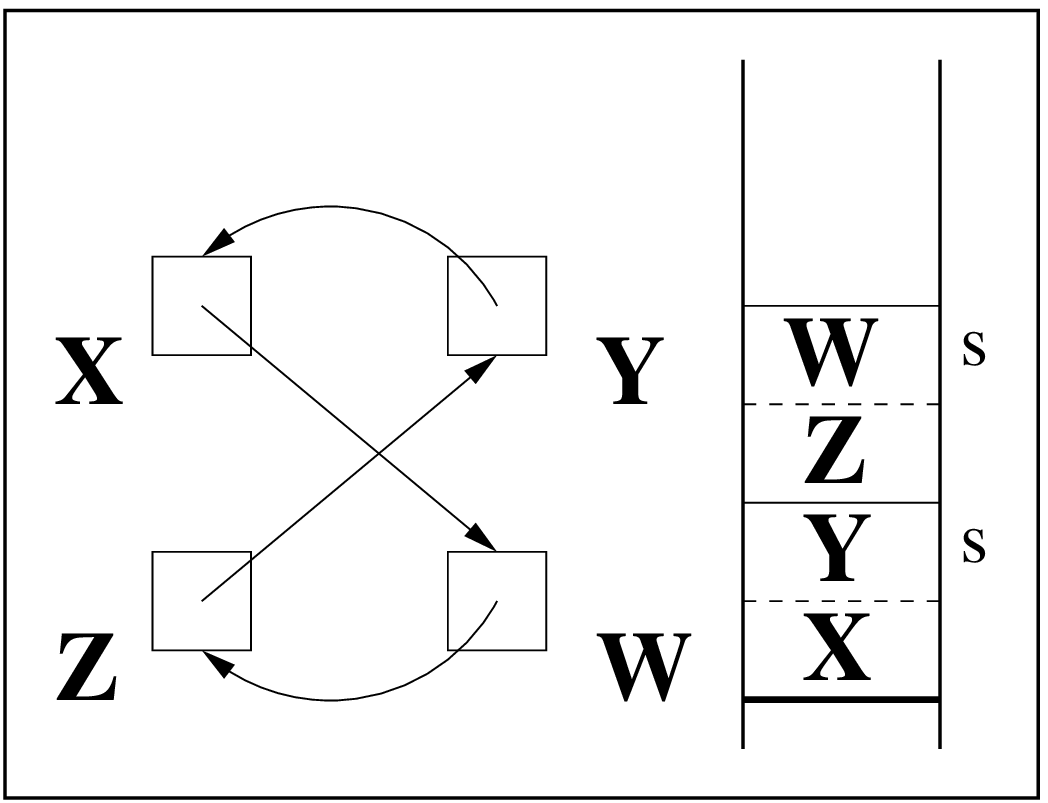,width=.20\textwidth}} \label{ex2step2}}
\subfigure[\texttt{Untrail}]{{\epsfig{file=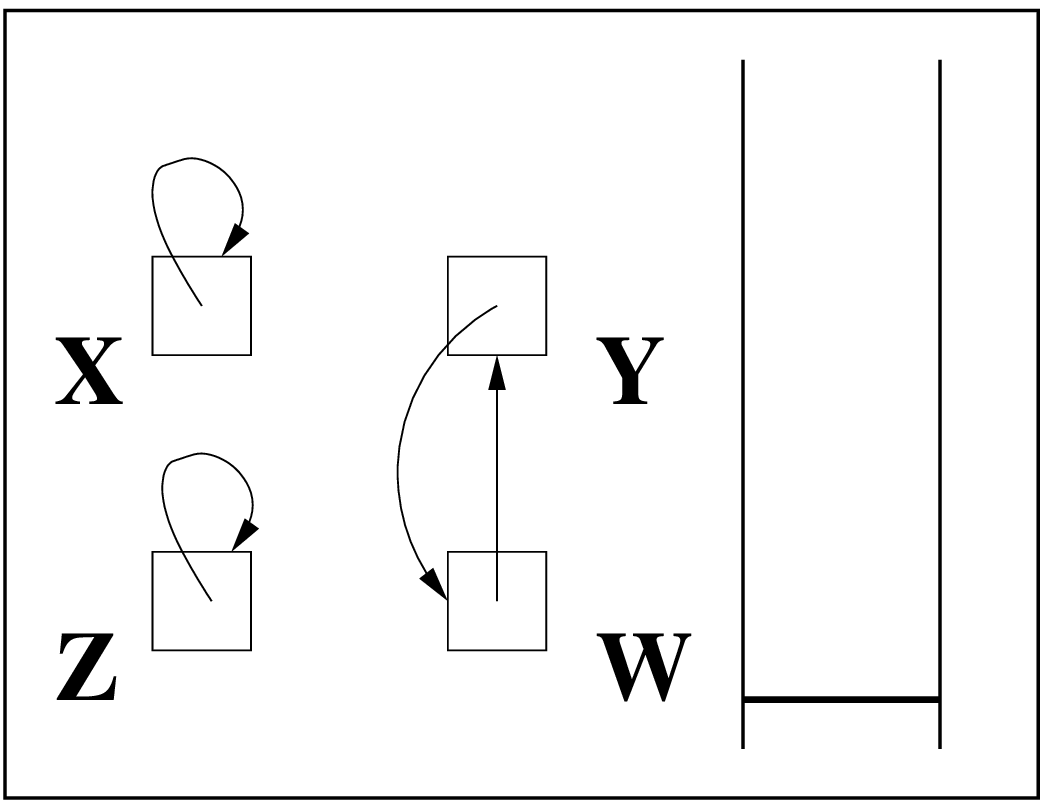,width=.20\textwidth}} \label{ex2untrail}}
\caption{Counterexample of incremental behavior of swap trailing: it does not
eliminate the need for further trailing of the same cells.}
\label{swaptrailfig}
\end{centering}
\end{figure}

Next \texttt{X} and \texttt{Z} are aliased, creating one large chain (see
Figure~\ref{ex2step2}). During this step \texttt{X} and \texttt{Z} are not
(swap) trailed since they have already been swap trailed after the most
recent choice point (and we are assuming this means trailing is not needed). Finally, the
execution fails and untrailing tries to restore the situation at the most
recent choice point. However, Figure~\ref{ex2untrail} shows that the
omission of the last swap trailing was invalid, as untrailing fails to
restore the correct situation. Thus, a cell involved in swap trailing still
needs trailing later in the same segment of the execution.
\end{cxmpl}

\subsection{The \tdom{} analysis domain}

The implications for the \tdom{} analysis domain are simple: it only needs to
distinguish between variables that do not have to 
be trailed again (deep trailed)
and those which have to (rest).  In other words, 
variables can only have one of two possible states 
at a particular program point: deep trailed or not trailed at all.
Hence, the type of elements of our \tdom{} domain will be ${\cal P}(Old_p)$.
The ordering $\sqsubseteq$ is simply $\supseteq$. 

All the operations we have defined for the \ntdom{} domain have to be adapted
to this simplification. 
This adaptation is rather straightforward: every description
$l$ in \tdom{} is treated as if it were the description $(\emptyset,l)$ in \ntdom{}, and new descriptions $l'$ in \tdom{} are obtained by first calculating the $(s',d')$ descriptions using the \ntdom{}
operations and then setting $l' = d'$.

\subsection{Optimization based on the analysis}\label{opts}

Again, the pre-description of every unification is used to improve that unification.
The possible optimizations based on the \tdom{} domain are more limited than
those for the \ntdom{} domain, as only deep trailed variables are represented
in the descriptions:

\begin{itemize}
\item
  For the unification of two variables, 
  a variant without (swap) trailing can be used
  if both variables are in the pre-description (i.e.~deep-trailed).
\item
  For the binding of an unbound variable $Y$ to a term $f(X_1,...,X_n)$,
  a variant of the unification without (chain) trailing can be used if $Y$
  is in the pre-description. In addition, no (swap) trailing is required
  for any of the $X_i$ that appear in the pre-description.
\item
  For the unification of two bound variables, if 
  both variables are in the pre-description,
  or if one is in the pre-description and the other is known to be ground, then no trailing
  is needed at runtime. This means that if 
  during the recursive unification process of
  the bound variables, unbound variables are 
  unified or bound, nothing will need to be trailed
  for these unbound variables.
\end{itemize}

\section{Experimental Results}\label{results}

We first examine the effect of the trailing analysis \ntdom{} and
its associated optimizations on the classic PARMA trailing scheme
for HAL.
We then look at the effect of the improved PARMA trailing scheme,
and at the effect of the use of the trailing analysis \tdom{} on the
improved PARMA trailing scheme.
Finally, we examine the  improved PARMA trailing scheme in the context
of dProlog.
All timing results were obtained on an Intel Pentium 4 2.00 GHz 512 MB.

\subsection{Effect of trailing analysis using \ntdom{} in HAL}
\label{results-class}

The \ntdom{} analyzer has been implemented in the analysis framework of HAL and
applied to six HAL benchmarks that use the Herbrand solver: \texttt{icomp},
\texttt{hanoi}, \texttt{qsort}, \texttt{serialize}, \texttt{warplan} and
\texttt{zebra}. Table \ref{tab:hal:benchmarks} gives a summary of these
benchmarks.
All benchmarks make use of the Herbrand solver 
and cannot be executed as Mercury programs (without significantly modifying the
algorithm and representation).

\begin{table}
\caption{HAL Benchmark descriptions and lines of code}\label{tab:hal:benchmarks}
\centerline{
\begin{tabular}{llr}
\hline\hline
Benchmark       & Description                                            & Lines\\
\cline{1-3}
\texttt{icomp}     & a cut down version of the interactive BIM compiler  & 294 \\
\texttt{hanoi}     & the Hanoi puzzle using difference lists             &  31 \\
\texttt{qsort}     & the quick sort algorithm using difference lists     &  43 \\
\texttt{serialize} & the classic Prolog palindrome benchmark             &  74 \\
\texttt{warplan}   & war planner for robot control                       & 316 \\
\texttt{zebra}     & the classic five houses puzzle                      &  82 \\
\hline\hline
\end{tabular}
}
\end{table}

The pre-descriptions inferred
for the unifications of these 
benchmarks have been used to optimize the generated Mercury
code by avoiding unnecessary trailing, as explained 
in Section~\ref{sec:ntdom:opt}.

\begin{table}
\caption{Compilation statistics for notrail analysis}\label{tab:ca:static}
\centerline{
\begin{tabular}{lrrrrrrrrr}
\hline\hline
Benchmark       & \multicolumn{3}{c}{Compilation Time}         & \multicolumn{3}{c}{Old unifications} & \multicolumn{1}{c}{Size} \\
                & Analysis      & Total         & Relative & Improved      & Total & Relative & Relative \\
\cline{1-8}
\texttt{icomp} & 1.170 & 2.110 & 55.5 \% & 314 & 1,542 & 20.4 \%& 120.5 \% \\
\texttt{hanoi} & 0.030 & 0.350 & 8.6 \% & 13 & 13 & 100.0 \%& 100.0 \% \\
\texttt{qsort} & 0.020 & 0.810 & 2.5 \% & 7 & 7 & 100.0 \%& 100.0 \% \\
\texttt{serialize} & 0.040 & 0.430 & 9.3 \% & 1 & 20 & 5.0 \%& 100.2 \% \\
\texttt{warplan} & 1.080 & 2.590 & 41.7 \% & 93 & 1,347 & 6.9 \%& 156.2 \% \\
\texttt{zebra} & 0.090 & 0.560 & 16.1 \% & 40 & 177 & 22.6 \%& 108.6 \% \\
\hline\hline
\end{tabular}
}
\end{table}

Table \ref{tab:ca:static} shows, for each benchmark, the analysis time in
seconds compared to the total compilation time, the number of improved
unifications compared to the total number of unifications involving old
variables, and the size of generated binary executable. The binary size of
the optimized program is expressed as the number of bytes relative to the
unoptimized program.

The high compilation times obtained for some benchmarks are due to the
existence of predicates with many different pre-descriptions, something the
analysis has not been optimized for yet. The deterministic nature of both
\texttt{hanoi} and \texttt{qsort} benchmarks, allows the analysis to infer
that all unifications should be replaced by a non-trailing alternative. In
the other benchmarks a much smaller fraction of unifications can be improved
due to the heavy use of non-deterministic predicates.

The last table shows that due to the multi-variant specialization, there
may be a considerable size blow-up. In particular, for \texttt{icomp} and
\texttt{warplan} the size is substantially increased.  Various approaches to limit
the number generated variants, explored in other work, apply to this work as
well. For example, one approach is to use profiling information to only retain
the most performance-critical variants (see \cite{FerreiraDamas}). Another
approach, taken in \cite{Nancy}, is to only generate the most and least
optimized variants. The latter would reproduce the optimal result for
\texttt{hanoi} and \texttt{qsort}.

\begin{table}
\caption{Benchmark timings for classic PARMA: unoptimized (cparma) and optimized with trailing analysis (caparma)}\label{tab:cca:timings}
\centerline{
\begin{tabular}{lrrrr}
\hline\hline
Benchmark       &       Iterations & \multicolumn{3}{c}{Time} \\   
                &                  & cparma & caparma & relative \\
\cline{1-5}
\texttt{icomp}           & 10,000           & 0.834 & 0.790 & 94.7 \% \\
\texttt{hanoi}           &     10           & 0.990 & 0.707 & 71.4 \% \\
\texttt{qsort}           & 10,000           & 0.363 & 0.303 & 83.5 \% \\
\texttt{serialize}       & 10,000           & 0.901 & 0.884 & 98.1 \% \\
\texttt{warplan}         &     10           & 1.293 & 1.407 & 108.8 \% \\
\texttt{zebra}           &    200           & 1.239 & 1.254 & 101.2 \% \\
\hline\hline
\end{tabular}
}
\end{table}

Table~\ref{tab:cca:timings} presents the execution times in seconds obtained by executing
each benchmark a number of times in a loop; the iteration
number in the table gives that loop count. This execution process (and the iteration number) is also used to obtain all other results 
shown for these HAL benchmarks.

The significant speed-up obtained for both the \texttt{hanoi} and \texttt{qsort} benchmarks is
explained by the effects of replacing all unifications with a non-trailing
version on the maximum size of the trail stack (in kilobytes), and on the
total number of trailing operations, as shown in Table \ref{tab:cca:trailings}.
In the non-deterministic benchmarks, a much smaller fraction of the trailing
operations is removed. This results in a smaller speed-up or even a slight
slow-down. The slow-down shows that the optimization does not come without
a cost.

\begin{table}
\caption{Benchmark trail sizes for classic PARMA: 
unoptimized (cparma) and optimized with trailing analysis (caparma)}\label{tab:cca:trailings}
\centerline{
\begin{tabular}{lrrrrrr}
\hline\hline
Benchmark       & \multicolumn{3}{c}{Maximum trail}   & \multicolumn{3}{c}{Trailing operations}    \\
                &    cparma & caparma  & relative &  cparma  & caparma & relative \\
\cline{1-7}\cline{1-7}
\texttt{icomp} & 5,545 & 4,217 & 76.1 \% & 1,110 & 860 & 77.5 \% \\
\texttt{hanoi} & 61,441 & 0 & 0.0 \% & 7,864,300 & 0 & 0.0 \% \\
\texttt{qsort} & 11,801 & 0 & 0.0 \% & 1,510 & 0 & 0.0 \% \\
\texttt{serialize} & 16,569 & 12,657 & 76.4 \% & 2,120 & 1,620 & 76.4 \% \\
\texttt{warplan} & 17 & 9 & 52.9 \% & 102,290 & 101,820 & 99.5 \% \\
\texttt{zebra} & 209 & 185 & 88.5 \% & 5,153,800 & 4,920,600 & 95.5 \% \\
\hline\hline
\end{tabular}
}
\end{table}

The larger active code size due to the multi-variant specialization
has an impact on the instruction cache behavior. Table~\ref{tab:cca:cache}
shows the impact on instruction references and instruction cache misses,
obtained with the cachegrind skin of the valgrind memory debugger (see
\cite{nethercote03valgrind}).  The number of instruction references is the
number of times an instruction is retrieved from memory and the instruction
cache miss rate is the percentage of instruction references in main memory
instead of cache.

The table clearly shows that the elimination of all trailing operations results
in a considerable reduction of executed instructions. On the other side of
the spectrum, the multi-variant specialization has a negative effect on the
instruction cache miss rate, 
which explains the slow-down of the \texttt{warplan} benchmark.

\begin{table}
\caption{Benchmark instruction cache misses for classic PARMA: unoptimized (cparma) vs.
optimized with trailing analysis (caparma)}\label{tab:cca:cache}
\centerline{
\begin{tabular}{lrrrrrr}
\hline\hline
Benchmark       & \multicolumn{3}{c}{I1 instruction cache miss rate} & \multicolumn{3}{c}{Instruction references}\\
                &    cparma & caparma & relative &    cparma & caparma & relative \\
\cline{1-7}
\texttt{icomp} & 0.85 \% & 1.79 \% & 210.6 \% & 716 $\times 10^6$ & 709 $\times 10^6$ & 99.0 \%  \\
\texttt{hanoi} & 0.00 \% & 0.00 \% & - \% & 991 $\times 10^6$ & 839 $\times 10^6$ & 84.7 \%  \\
\texttt{qsort} & 0.00 \% & 0.00 \% & - \% & 427 $\times 10^6$ & 397 $\times 10^6$ & 93.0 \%  \\
\texttt{serialize} & 0.00 \% & 0.70 \% & $\infty$ \% & 912 $\times 10^6$ & 899 $\times 10^6$ & 98.6 \%  \\
\texttt{warplan} & 1.55 \% & 4.44 \% & 286.5 \% & 1,559 $\times 10^6$ & 1,560 $\times 10^6$ & 100.1 \%  \\
\texttt{zebra} & 0.40 \% & 0.10 \% & 25.0 \% & 1,300 $\times 10^6$ & 1,291 $\times 10^6$ & 99.3 \%  \\
\hline\hline
\end{tabular}
}
\end{table}

\subsection{Effect of the improved trailing scheme in the Mercury back-end of HAL}

The improved unconditional PARMA 
trailing scheme has also been implemented in the
Mercury back-end of HAL. 
Since Mercury already has a tagged trail, this was not too difficult.
Aside from the discussed trailings for unification,
this system also requires trailing when a term is constructed with an old
variable as an argument. In this term construction, the argument cell
in the term structure is inserted in the variable chain. This modifies
one cell in the old variable chain.  In the classic scheme this cell is
trailed with value trailing.  To avoid value trailing altogether this
has been replaced with swap trailing in the improved trailing scheme.

Table \ref{tab:ci:timingstrailings} presents the timing and maximal trail for both the
classic and improved trailing scheme for the six HAL benchmarks used before.

\setlength{\tabcolsep}{3pt}
\begin{table}[h]
\caption{Timing and maximal trail for the classic (cparma) and improved
(iparma) unconditional PARMA trailing scheme for the Mercury back-end of
  HAL.}\label{tab:ci:timingstrailings}
\begin{center}
\begin{tabular}{lrrrrrr}
\hline\hline
Benchmark       &       \multicolumn{3}{c}{Time}              &       \multicolumn{3}{c}{Maximal trail}      \\
                & cparma        &iparma  & relative & cparma        & iparma & relative       \\
\cline{1-7}
\texttt{icomp} & 0.834 & 0.809 & 97.0 \% & 5,545 & 3,049 & 55.0 \% \\
\texttt{hanoi} & 0.990 & 0.944 & 95.4 \% & 61,441 & 40,961 & 66.7 \% \\
\texttt{qsort} & 0.363 & 0.350 & 96.4 \% & 11,801 & 7,857 & 66.6 \% \\
\texttt{serialize} & 0.901 & 0.836 & 92.8 \% & 16,569 & 10,233 & 61.8 \% \\
\texttt{warplan} & 1.293 & 1.284 & 99.3 \% & 17 & 9 & 52.9 \% \\
\texttt{zebra} & 1.239 & 1.171 & 94.5 \% & 209 & 105 & 50.2 \% \\
\hline\hline
\end{tabular}
\end{center}
\end{table}

In all benchmarks the improved trailing scheme is faster than the
classic scheme. The differences
are a few percentages though, 
with a maximum difference of slightly more than 7\% for
the \texttt{serialize} benchmark. Much more important are the effects of the improved
trailing scheme on the maximal trail size. The maximal trail is at
least 30\% and up to 50\% smaller for the improved scheme than for the
classic scheme.

\subsection{Effect of the improved trailing scheme combined with 
trailing analysis \tdom{} in the Mercury back-end of HAL}

The trailing analysis presented in Section~\ref{analysis} and implemented in
HAL, was modified, as proposed in Section~\ref{analysis-impr}, to deal with
the improved trailing scheme.
Table \ref{tab:iia:timingstrailings} presents the timing and maximal trail
for the HAL benchmarks obtained under the improved scheme with the
information inferred by the modified analysis, and compares the results
obtained under the same scheme without any analysis information.

\begin{table}[h]
\caption{Timing and maximal trail for the improved unconditional PARMA scheme
without (iparma) and with (iaparma) 
\tdom{} trailing analysis, relative to the classic scheme without trailing.}\label{tab:iia:timingstrailings}
\begin{center}
\begin{tabular}{lrrrrrr}
\hline\hline
Benchmark       & \multicolumn{3}{c}{Time}    & \multicolumn{3}{c}{Maximal trail}      \\
                & iparma    & iaparma & relative       & iparma        & iaparma & relative\\
\cline{1-7}
\texttt{icomp} & 97.0 \% & 93.3 \% & 96.2 \% & 55.0 \%& 47.9 \%& 87.1 \% \\
\texttt{hanoi} & 95.4 \% & 71.6 \% & 75.1 \% & 66.7 \%& 0.0 \%& 0.0 \% \\
\texttt{qsort} & 96.4 \% & 83.5 \% & 86.6 \% & 66.6 \%& 0.0 \%& 0.0 \% \\
\texttt{serialize} & 92.8 \% & 92.8 \% & 100.0 \% & 61.8 \%& 61.8 \%& 100.0 \% \\
\texttt{warplan} & 99.3 \% & 99.7 \% & 100.4 \% & 52.9 \%& 52.9 \%& 100.0 \% \\
\texttt{zebra} & 94.5 \% & 91.9 \% & 97.3 \% & 50.2 \%& 46.4 \%& 92.4 \% \\
\hline\hline
\end{tabular}
\end{center}
\end{table}

For the \texttt{serialize} and \texttt{warplan} 
benchmarks the analysis was not able to
reduce the number of actual trailing operations.
For the other four benchmarks the combination of the improved scheme with
analysis yields better results, both for time and maximal trail. For the 
\texttt{hanoi}
and \texttt{qsort}  benchmarks 
there is again a drastic improvement: all trailings have been
avoided, with a distinctive time improvement of 25\% and 15 \% respectively. For the other two
benchmarks, \texttt{icomp} and \texttt{zebra}, 
there is a maximal trail improvement of about
10\% together with a slightly reduced time, 4\% and 3\% better respectively.
Overall, the combination of the improved scheme with the trailing analysis
never makes the results worse. Since it drastically improves some
benchmarks and shows a modest improvement of others, it is fair to
conclude that the combination is superior to the improved system
without analysis.

\subsection{Effect of the improved trailing scheme in dProlog}

Let us now present the experimental results of the improved conditional PARMA
trailing scheme in dProlog for several small benchmarks and one bigger
program, \texttt{comp}. 
Table \ref{tab:dprolog} shows the timing and maximal trail use for each
benchmark. Time is given in seconds and applies to the number of runs (iterations) given.
The maximal trail size is given in kilobytes and applies to a
single run.
 
\setlength{\tabcolsep}{3pt}
\begin{table}[h]
\caption{PARMA in dProlog: classic (cparma) vs. improved trailing (iparma)}\label{tab:dprolog}
\begin{center}
\begin{tabular}{lrrrrr}
\hline\hline
Benchmark       &      Iterations  &   \multicolumn{2}{c}{Time}      &       \multicolumn{2}{c}{Maximal trail}      \\
                & & cparma        & iparma                &       cparma          & iparma        \\
\cline{1-6}
\texttt{boyer} &10&  		.950&    .920& 450.6 &225.3 \\
\texttt{browse} &10& 		1.010&   1.010&5.2 &4.5 \\
\texttt{cal} &100 &    		1.800&   1.800&0.4 &0.2 \\
\texttt{chat} &50 &  		1.020&   1.040&3.6 &1.9 \\
\texttt{crypt} &2,000 & 	1.160&   1.170&0.5 &0.2 \\
\texttt{ham} &20 &    		1.160&   1.130&0.8 &0.4 \\
\texttt{meta\_qsort} &1,250 &   1.070&   1.090&12.6 &7.4 \\
\texttt{nrev} &50,000 &   	.900&    .860& 0.4 &0.2 \\
\texttt{poly\_10} &100 &       	.630&    .650& 52.6 &26.3 \\
\texttt{queens\_16} &20 &     	1.810&   1.790&0.7 &0.3 \\
\texttt{queens} &100 & 		3.310&   3.300&0.7 &0.3 \\
\texttt{reducer} &200 &        	.440&    .430& 18.9 &10.0 \\
\texttt{sdda} &12,000 &   	1.000&   1.010&1.3 &0.8 \\
\texttt{send} &100 &   		.800&    .800& 0.5 &0.2 \\
\texttt{tak} &100 &    		1.620&   1.520&373.1 &186.6 \\
\texttt{zebra} &300 &  		2.510&   2.730&1.6 &0.8 \\

\cline{1-6}                                                                             
\multicolumn{2}{c}{relative average}&	100\%&	 99.9\%&   100\%       &     51.7\%    \\
\cline{1-6}
comp         &       1       & 1.930   &  1.890                &  2516.3 &1319.8 \\
\cline{1-6}
\multicolumn{2}{c}{comp relative}	&	100\%	&  97.9\%		&  100\%		& 52.4\%	\\
\hline\hline
\end{tabular}
\end{center}
\end{table}

The time difference between the classic and the improved scheme is negligible.
The improved scheme is at most 8.8\% slower, for the \texttt{zebra} 
benchmark, but on
average both are about equally fast. The price for the lower trail usage is an
increase in instructions executed and that is why there is no net speedup.

The differences in maximal trail use however are substantial. While swap
trail and chain trail halve the trail stack consumption, value trailing is
still used for some cases of variable--variable trailing. Yet experimental
results show that that kind of variable--variable trailing does not occur very
often in most benchmarks, as the maximal trail stack is effectively halved
in eleven benchmarks and on average the maximal trail use is 51.7\% of
the classical scheme.

The results for the smaller benchmarks are confirmed by the larger 
\texttt{comp} program. 
Execution time is nearly the same for the classic and improved
trailing scheme and the maximal trail shows a similar improvement of almost
50\%.

\section{Related and future work}\label{future}

As far as
we know, the modifications suggested to 
the classic PARMA trailing scheme are new.

A somewhat similar analysis for detecting variables that do not have
to be trailed is presented by Debray in \cite{debray} together with
corresponding optimizations.  Debray's analysis however is for the WAM
variable representation and in a traditional Prolog setting, i.e.,
without type, mode and determinism declarations. Also in
\cite{Aquarius} trailing is avoided, but only for variables that are
{\em new} in our terminology and, again, the setting is basically the
WAM representation.

Taylor too keeps track of a trailing state of variables in the global
analysis of his PARMA system with the classic PARMA trailing scheme
(see \cite{taylorthesis,taylor89}). As opposed to the \ntdom{} 
analysis we have presented here, Taylor's analysis is less
precise and closer to the \tdom{} analysis presented here: the
trailing state of a variable can only be that it has to be trailed or
not, i.e. there is no intermediary shallow trailing state.
 
There exist also two run-time technique for preventing the multiple
value trailing between two choice points.  The first, described in
\cite{jacquesnoyephd}, only works in the WAM scheme, because it introduces
linear reference chains that PARMA does not allow.  The second, described in
\cite{beldiceanu-timestamp}, maintains a timestamp for every cell that
corresponds to the choicepoint before the last update. However,
such a timestamp requires additional space, even in the case that the cell
is never updated. In the context of PARMA, timestamps would likely
consume more space than is actually saved by avoiding trailing.

Finally, there are other approaches
to the reconstruction of state on backtracking other than trailing,
using either 
copying~\cite{Schulte:99a} or recomputation~\cite{backtracknotrail}.
While PARMA (and for that matter WAM) bindings do not keep
enough information to allow recomputation on backtracking,
a copying approach to backtracking in PARMA is quite feasible.
This remains as an interesting question for future work.

There is little room left for optimization of the trailing analysis for the
improved unconditional trailing scheme. Of course, the analysis itself can be
improved by adopting a more refined representation for bound variables. Currently, all
PARMA chains in the structure of a bound variable are represented by the same
trailing state. Bound variables could be represented more accurately, by
requiring the domain to keep track of the different chains contained in the
structures to which the program variables are bound, their individual trailing
state and how these are affected by the different program constructs. Known
techniques (see for instance 
\cite{gerda,pascal,mulkers,lagoontypeframe,lagooniclp03}) based on type
information could be used to keep track of the constructor that a variable is
bound to and the trailing state of the different arguments, thereby making this
approach possible. This applies equally to the analysis of the classical
scheme.

Additionally, it would be interesting to see how much extra gain analysis can
add to the improved conditional trailing scheme as implemented in dProlog or in
the Mercury back-end of HAL that supports conditional trailing.
Such analysis would certainly not improve the maximal trail, but it would
remove the overhead of the run-time test. This will most likely 
also result in a
small speed-up.

Though experimental results show that the improved scheme with analysis is
better than the classic scheme with analysis, this need not be true for all
programs. Recall that between two choice points all value trailings of a cell
but the first can be eliminated in the classic scheme, while no swap trailings
could be eliminated in the improved scheme. A hybrid scheme would 
be possible using analysis to decide 
on a single unification basis if either swap trailing or value
trailing is better at minimizing the amount of trailing and the cost of
untrailing. This analysis would require a more global view of all the trailings
in between two choice points. Moreover, some trailings could be common to
different pairs of choice points and optimality would depend on where execution
spends most of its time.

Also the untrailing operation can be improved: when analysis is able to
determine for instance that the only trailing that happened was a swap
trailing, no tags need to be set and tested.
\vspace{-1mm}
\section*{Acknowledgements}
\vspace{-1mm}
We would like to thank the referees for their detailed and insightful
reports which have significantly improved the paper. \vspace{-3mm}


\bibliography{trailing}

\begin{thebibliography}{}

\bibitem[\protect\citeauthoryear{Aggoun and Beldiceanu}{Aggoun and
  Beldiceanu}{1990}]{beldiceanu-timestamp}
{\sc Aggoun, A.} {\sc and} {\sc Beldiceanu, N.} 1990.
\newblock {Time Stamps Techniques for the Trailed Data in Constraint Logic
  Programming Systems}.
\newblock In {\em SPLT'90: 8$^{\mbox{{\`e}me}}$ S{\'e}minaire Programmation en
  Logique}, {S.~Bourgault} {and} {M.~Dincbas}, Eds. CNET, Tr{\'e}gastel,
  France, 487--510.

\bibitem[\protect\citeauthoryear{A{\"\i}t-Kaci}{A{\"\i}t-Kaci}{1991}]{ait-kaci%
91wam}
{\sc A{\"\i}t-Kaci, H.} 1991.
\newblock {\em {Warren's Abstract Machine: A Tutorial Reconstruction}}.
\newblock {MIT} {P}ress.

\bibitem[\protect\citeauthoryear{Bruynooghe}{Bruynooghe}{1991}]{bruy91}
{\sc Bruynooghe, M.} 1991.
\newblock {A Practical Framework for the Abstract Interpretation of Logic
  Programs}.
\newblock {\em Journal of Logic Programming\/}~{\em 10,\/}~1/2/3{\&}4, 91--124.

\bibitem[\protect\citeauthoryear{Bueno, de~la Banda, Hermenegildo, Marriott,
  Puebla, and Stuckey}{Bueno et~al\mbox{.}}{2001}]{modular-anal-lopstr-formal}
{\sc Bueno, F.}, {\sc de~la Banda, M. J.~G.}, {\sc Hermenegildo, M.~V.}, {\sc
  Marriott, K.}, {\sc Puebla, G.}, {\sc and} {\sc Stuckey, P.~J.} 2001.
\newblock {A Model for Inter-module Analysis and Optimizing Compilation}.
\newblock In {\em LOPSTR '00: Selected Papers form the 10th International
  Workshop on Logic Based Program Synthesis and Transformation}, {K.-K. Lau},
  Ed. Lecture Notes in Computer Science, vol. 2042. Springer Verlag, London,
  UK, 86--102.

\bibitem[\protect\citeauthoryear{de~la Banda, Demoen, Marriott, and
  Stuckey}{de~la Banda et~al\mbox{.}}{2002}]{flops2002}
{\sc de~la Banda, M. J.~G.}, {\sc Demoen, B.}, {\sc Marriott, K.}, {\sc and}
  {\sc Stuckey, P.~J.} 2002.
\newblock {To the Gates of HAL: A HAL Tutorial}.
\newblock In {\em FLOPS 2002: Proceedings of the 6th International Symposium on
  Functional and Logic Programming}, {Z.~Hu} {and}
  {M.~Rodr\'{\i}guez-Artalejo}, Eds. Lecture Notes in Computer Science, vol.
  2441. Springer Verlag, Aizu, Japan, 47--66.

\bibitem[\protect\citeauthoryear{de~la Banda, Marriott, Stuckey, and
  S{\o}ndergaard}{de~la Banda et~al\mbox{.}}{1998}]{diff}
{\sc de~la Banda, M. J.~G.}, {\sc Marriott, K.}, {\sc Stuckey, P.~J.}, {\sc
  and} {\sc S{\o}ndergaard, H.} 1998.
\newblock {Differential Methods in Logic Program Analysis}.
\newblock {\em Journal of Logic Programming\/}~{\em 35,\/}~1, 1--37.

\bibitem[\protect\citeauthoryear{Debray}{Debray}{1992}]{debray}
{\sc Debray, S.} 1992.
\newblock {A Simple Code Improvement Scheme for Prolog}.
\newblock {\em Journal of Logic Programming\/}~{\em 13,\/}~1 (May), 57--88.

\bibitem[\protect\citeauthoryear{Demoen, de~la Banda, Harvey, Marriott, and
  Stuckey}{Demoen et~al\mbox{.}}{1999}]{demoen99overview}
{\sc Demoen, B.}, {\sc de~la Banda, M. J.~G.}, {\sc Harvey, W.}, {\sc Marriott,
  K.}, {\sc and} {\sc Stuckey, P.~J.} 1999.
\newblock {An Overview of HAL}.
\newblock In {\em CP'99: Proceedings of the 5th International Conference on
  Principles and Practice of Constraint Programming}, {J.~Jaffar}, Ed. Lecture
  Notes in Computer Science, vol. 1713. Springer Verlag, Alexandria, Virginia,
  USA, 174--188.

\bibitem[\protect\citeauthoryear{Demoen and Nguyen}{Demoen and
  Nguyen}{2000}]{dProlog}
{\sc Demoen, B.} {\sc and} {\sc Nguyen, P.-L.} 2000.
\newblock {{S}o Many {WAM} Variations, so Little Time}.
\newblock In {\em CL2000: Proceedings of the 1st International Conference on
  Computational Logic}, {J.~Lloyd}, {V.~Dahl}, {U.~Furbach}, {M.~Kerber},
  {K.-K. Lau}, {C.~Palamidessi}, {L.~Moniz~Pereira}, {Y.~Sagiv}, {and}
  {P.~J.~Stuckey}, Eds. Lecture Notes in Artificial Intelligence, vol. 1861.
  ALP, Springer Verlag, London, UK, 1240--1254.

\bibitem[\protect\citeauthoryear{Ferreira and Damas}{Ferreira and
  Damas}{2003}]{FerreiraDamas}
{\sc Ferreira, M.} {\sc and} {\sc Damas, L.} 2003.
\newblock {Controlling Code Expansion in a Multiple Specialization Prolog
  Compiler}.
\newblock In {\em {P}roceedings of {CICLOPS} 2003: {C}olloquium on
  {I}mplementation of {C}onstraint and {LO}gic {P}rogramming {S}ystems},
  {R.~Lopes} {and} {M.~Ferreira}, Eds. University of Porto, Mumbai, India,
  75--87.
\newblock Technical Report DCC-2003-05, DCC - FC \& LIACC, Univeristy of Porto,
  December 2003.

\bibitem[\protect\citeauthoryear{Janssens and Bruynooghe}{Janssens and
  Bruynooghe}{1993}]{gerda}
{\sc Janssens, G.} {\sc and} {\sc Bruynooghe, M.} 1993.
\newblock {Deriving Descriptions of Possible Value of Program Variables by
  means of Abstract Interpretation}.
\newblock {\em Journal of Logic Programming\/}~{\em 13}, 205--258.

\bibitem[\protect\citeauthoryear{Lagoon, Mesnard, and Stuckey}{Lagoon
  et~al\mbox{.}}{2003}]{lagooniclp03}
{\sc Lagoon, V.}, {\sc Mesnard, F.}, {\sc and} {\sc Stuckey, P.~J.} 2003.
\newblock {Termination Analysis with Types Is More Accurate}.
\newblock In {\em ICLP 2003: Proceedings of the 19th International Conference
  on Logic Programming}, {C.~Palamidessi}, Ed. Lecture Notes in Computer
  Science, vol. 2916. Springer Verlag, Mumbai, India, 254--268.

\bibitem[\protect\citeauthoryear{Lagoon and Stuckey}{Lagoon and
  Stuckey}{2001}]{lagoontypeframe}
{\sc Lagoon, V.} {\sc and} {\sc Stuckey, P.} 2001.
\newblock {A Framework for Analysis of Typed Logic Programs}.
\newblock In {\em FLOPS 2001: Proceedings of the 5th International Symposium on
  Functional and Logic Programming}, {H.~Kuchen} {and} {K.~Ueda}, Eds. Lecture
  Notes in Computer Science, vol. 2024. Springer Verlag, Tokyo, Japan,
  296--310.

\bibitem[\protect\citeauthoryear{Lindgren, Mildner, and Bevemyr}{Lindgren
  et~al\mbox{.}}{1995}]{lindgren95taylor}
{\sc Lindgren, T.}, {\sc Mildner, P.}, {\sc and} {\sc Bevemyr, J.} 1995.
\newblock {On Taylor's Scheme for Unbound Variables}.
\newblock Tech. rep., Computer Science Department, Uppsala University. Oct.

\bibitem[\protect\citeauthoryear{Mazur}{Mazur}{2001}]{Nancy}
{\sc Mazur, N.} 2001.
\newblock {Compile-time Garbage Collection for the Declarative Language
  Mercury}.
\newblock Ph.D. thesis, Department of Computer Science, K.U.Leuven, Leuven,
  Belgium.

\bibitem[\protect\citeauthoryear{Mulkers, Winsborough, and Bruynooghe}{Mulkers
  et~al\mbox{.}}{1994}]{mulkers}
{\sc Mulkers, A.}, {\sc Winsborough, W.~H.}, {\sc and} {\sc Bruynooghe, M.}
  1994.
\newblock {Live-Structure Dataflow Analysis for Prolog}.
\newblock {\em ACM Transactions on Programming Languages and Systems\/}~{\em
  16,\/}~2, 205--258.

\bibitem[\protect\citeauthoryear{Nethercote}{Nethercote}{2001}]{nethercote01th%
esis}
{\sc Nethercote, N.} 2001.
\newblock {T}he {A}nalysis {F}ramework of {HAL}.
\newblock M.S.\ thesis, University of Melbourne.

\bibitem[\protect\citeauthoryear{Nethercote and Seward}{Nethercote and
  Seward}{2003}]{nethercote03valgrind}
{\sc Nethercote, N.} {\sc and} {\sc Seward, J.} 2003.
\newblock {Valgrind: A Program Supervision Framework}.
\newblock {\em Electronic Notes in Theoretical Computer Science\/}~{\em
  89,\/}~2, 1--23.

\bibitem[\protect\citeauthoryear{Noy\'e}{Noy\'e}{1994}]{jacquesnoyephd}
{\sc Noy\'e, J.} 1994.
\newblock {Elagage de contexte, retour arri\`ere superficiel, modifications
  r\'eversibles et autres: une \'etude approfondie de la WAM}.
\newblock Ph.D. thesis, Universit\'e de Rennes I.

\bibitem[\protect\citeauthoryear{Schulte}{Schulte}{1999}]{Schulte:99a}
{\sc Schulte, C.} 1999.
\newblock {Comparing Trailing and Copying for Constraint Programming}.
\newblock In {\em Proceedings of the Sixteenth International Conference on
  Logic Programming}, {D.~\protect{De} Schreye}, Ed. MIT Press, Las Cruces, NM,
  USA, 275--289.

\bibitem[\protect\citeauthoryear{Somogyi, Henderson, and Conway}{Somogyi
  et~al\mbox{.}}{1996}]{somogyi95mercury}
{\sc Somogyi, Z.}, {\sc Henderson, F.}, {\sc and} {\sc Conway, T.} 1996.
\newblock {The Execution Algorithm of Mercury, an Efficient Purely Declarative
  Logic Programming Language}.
\newblock {\em Journal of Logic Programming\/}~{\em 29,\/}~1--3, 17--64.

\bibitem[\protect\citeauthoryear{S{\o}ndergaard}{S{\o}ndergaard}{1986}]{sonder%
86}
{\sc S{\o}ndergaard, H.} 1986.
\newblock {An Application of Abstract Interpretation of Logic Programs: Occur
  Check Reduction}.
\newblock In {\em ESOP 86: Proceedings of the European Symposium on
  Programming}, {B.~Robinet} {and} {R.~Wilhelm}, Eds. Lecture Notes in Computer
  Science, vol. 213. Springer Verlag, Saarbr{\"u}cken, Germany, 327--338.

\bibitem[\protect\citeauthoryear{Taylor}{Taylor}{1989}]{taylor89}
{\sc Taylor, A.} 1989.
\newblock {Removal of Dereferencing and Trailing in Prolog Compilation}.
\newblock In {\em {Proceedings of the 6th Internation Conference on Logic
  Programming}}, {G.~Levi} {and} {M.~Martelli}, Eds. {MIT Press}, Lisbon,
  Portugal, 48--60.

\bibitem[\protect\citeauthoryear{Taylor}{Taylor}{1991}]{taylorthesis}
{\sc Taylor, A.} 1991.
\newblock {High Performace Prolog Implementation}.
\newblock Ph.D. thesis, Basser Department of Computer Science.

\bibitem[\protect\citeauthoryear{Taylor}{Taylor}{1996}]{taylor96parma}
{\sc Taylor, A.} 1996.
\newblock {Parma - Bridging the Performance GAP Between Imperative and Logic
  Programming}.
\newblock {\em Journal of Logic Programming\/}~{\em 29,\/}~1-3, 5--16.

\bibitem[\protect\citeauthoryear{{Van Hentenryck}, Cortesi, and Charlier}{{Van
  Hentenryck} et~al\mbox{.}}{1995}]{pascal}
{\sc {Van Hentenryck}, P.}, {\sc Cortesi, A.}, {\sc and} {\sc Charlier, B.~L.}
  1995.
\newblock {Type analysis of Prolog using Type Graphs}.
\newblock {\em {Journal of Logic Programming}\/}~{\em 22,\/}~3, 179--209.

\bibitem[\protect\citeauthoryear{{Van Hentenryck} and Ramachandran}{{Van
  Hentenryck} and Ramachandran}{1995}]{backtracknotrail}
{\sc {Van Hentenryck}, P.} {\sc and} {\sc Ramachandran, V.} 1995.
\newblock {Backtracking without Trailing in CLP(Rlin)}.
\newblock {\em ACM Transactions on Programming Languages and Systems\/}~{\em
  17,\/}~4 (July), 635--671.

\bibitem[\protect\citeauthoryear{Van~Roy and Despain}{Van~Roy and
  Despain}{1992}]{Aquarius}
{\sc Van~Roy, P.} {\sc and} {\sc Despain, A.} 1992.
\newblock {High-Performance Logic Programming with the Aquarius Prolog
  Compiler}.
\newblock {\em IEEE Computer\/}~{\em 25,\/}~1, 54--68.

\end{thebibliography}

\end{document}